\documentclass[utf8]{frontiersFPHY} 

\setcitestyle{square} 
\usepackage{url,hyperref,microtype,subcaption}
\usepackage[onehalfspacing]{setspace}
\usepackage{natbib}
\usepackage{multirow}
\usepackage{pdflscape}
\usepackage{array}
\usepackage{tablefootnote}
\usepackage[flushleft]{threeparttable}

\def\keyFont{\fontsize{8}{11}\helveticabold }
\def\firstAuthorLast{Dumbovi\'c {et~al.}} 
\def\Authors{
Mateja Dumbovi\'c\,$^{1,*}$,
Ja\v{s}a \v{C}alogovi\'{c}\,$^{1}$,
Karmen Martini\'c\,$^{1}$,
Bojan Vr\v{s}nak\,$^{1}$,
Davor Sudar\,$^{1}$,
Manuela Temmer\,$^{2}$, and
Astrid Veronig\,$^{2,3}$}
\def\Address{
$^{1}$Hvar Observatory, Faculty of Geodesy, University of Zagreb, Croatia \\
$^{2}$Institute of Physics, University of Graz, Austria\\
$^{3}$Kanzelh\"ohe Observatory for Solar and Environmental Research, University of Graz, Austria }
\def\corrAuthor{Mateja Dumbovi\'c, Hvar Observatory, Faculty of Geodesy, University of Zagreb, Ka\v{c}i\'ceva 26, HR-10000, Zagreb, Croatia}
\def\corrEmail{mdumbovic@geof.hr}

\begin{document}
\onecolumn
\firstpage{1}

\title[Drag based model (DBM) tools]{Drag-based model (DBM) tools for forecast of coronal mass ejection arrival time and speed} 

\author[\firstAuthorLast ]{\Authors} 
\address{} 
\correspondance{} 

\extraAuth{}

\maketitle

\begin{abstract}
\section{}
Forecasting the arrival time of coronal mass ejections (CMEs) and their associated shocks is one of the key aspects of space weather research. One of the commonly used models is, due to its simplicity and calculation speed, the analytical drag-based model (DBM) for heliospheric propagation of CMEs. DBM relies on the observational fact that slow CMEs accelerate whereas fast CMEs decelerate, and is based on the concept of MHD drag, which acts to adjust the CME speed to the ambient solar wind. Although physically DBM is applicable only to the CME magnetic structure, it is often used as a proxy for the shock arrival. In recent years, the DBM equation has been used in many studies to describe the propagation of CMEs and shocks with different geometries and assumptions. Here we give an overview of the five DBM versions currently available and their respective tools, developed at Hvar Observatory and frequently used by researchers and forecasters. These include: 1) basic 1D DBM, a 1D model describing the propagation of a single point (i.e. the apex of the CME) or concentric arc (where all points propagate identically); 2) advanced 2D self-similar cone DBM, a 2D model which combines basic DBM and cone geometry describing the propagation of the CME leading edge which evolves self-similarly; 3) 2D flattening cone DBM, a 2D model which combines basic DBM and cone geometry describing the propagation of the CME leading edge which does not evolve self-similarly; 4) DBEM, an ensemble version of the 2D flattening cone DBM which uses CME ensembles as an input and 5) DBEMv3, an ensemble version of the 2D flattening cone DBM which creates CME ensembles based on the input uncertainties. All five versions have been tested and published in recent years and are available online or upon request. We provide an overview of these five tools, of their similarities and differences, as well as discuss and demonstrate their application.
\tiny
 \keyFont{ \section{Keywords:} coronal mass ejections, solar wind, interplanetary shocks, magnetohydrodynamical drag, space weather forecast}
\end{abstract}
\section{Introduction}
\label{intro}

Coronal mass ejections (CMEs) are one of most prominent drivers of space weather in the heliosphere. They are the causes of largest geomagnetic storms \citep[e.g.][]{zhang03} as they may carry enhanced and specifically oriented magnetic fields \citep[see e.g.][]{bothmer98,demoulin08}. Forecasting the arrival time of CMEs and their associated shocks is therefore one of the key aspects of space weather research. Therefore, there is a diversity of CME models available today, some focusing on the arrival time forecast only and other, more complex models include forecast of other CME properties as well \citep[see e.g.][and references therein]{siscoe06,zhao14,vourlidas19,zhang20}.

Propagation of CMEs in the heliosphere with the purpose of obtaining the CME time of arrival (ToA) and speed of arrival (SoA), can be modelled by empirical models \citep[e.g.][]{gopalswamy01,paouris17}, kinematic shock propagation models \citep[e.g.][]{dryer01,zhao16a,takahashi17}, machine-learning models \citep[e.g.][]{sudar16,liu18}, numerical 3D MHD models (e.g. H3DMHD model by \citet{wu11}, WSA-ENLIL+Cone model by \citet{odstrcil04}, EUHFORIA model by \citet{pomoell18}, CORHEL model by \citet{mikic99} or AWSoM model by \citet{holst14}) and drag-based models (see below). All CME propagation models need CME input as well as input of characteristics of the background solar wind, where both may have large uncertainties. Therefore, it is not surprising that despite their differences, ToA errors of different propagation models revolve around 10 hours \citep{riley18,vourlidas19}.

One of the most popular CME propagation setups used in forecast models today is the drag-based propagation. In this concept, the CME which is initially under the influence of Lorentz force, gravity and drag-force due to interaction with the ambient medium, at a certain distance from the Sun is influenced dominantly by the drag force \citep[see e.g.][and references therein]{zhang06,temmer16}. This concept is supported by the observational fact that slow CMEs accelerate whereas fast CMEs decelerate \citep{sheeley99, gopalswamy00,sachdeva15}. The drag force can be represented by the aerodynamic drag equation describing the kinetic drag effect in a fluid \citep{cargill04,vrsnak07}, however it should be noted that in the IP space, i.e. collisionless solar wind environment, the drag is caused primarily by the emission of MHD waves and not particle collisions \citep{cargill96}.

Drag-based models typically use the same form of the basic drag equation applied to various geometries representing the CME structure of different dimensionality, e.g. 1D Drag-Based Model \citep[DBM,][]{vrsnak13,vrsnak14} and Enhanced DBM \citep{hess14,hess15}, 2D Drag-Based Model \citep{zic15}, the 2D Ellipse Evolution Model \citep[ElEvo,][]{mostl15} and a version of ElEvo using data from Heliospheric Imagers \citep[ElEvoHi,][]{rollett16}, and 3D flux rope models such as ANother Type of Ensemble Arrival Time Results \citep[ANTEATR,][]{kay18} or 3-Dimensional Coronal ROpe Ejection \citep[3DCORE,][]{mostl18}. Since drag-based models use an analytical equation to describe the time-dependent evolution of the CME, they are computationally efficient and thus widely used in probabilistic/ensemble modelling approaches \citep[e.g.][]{napoletano18,dumbovic18a,amerstorfer18,kay18,kay20,amerstorfer20}. The advantage of ensemble modelling is that it gives the probability of arrival, as well as the range of possible arrival times and speeds.

Starting with a basic 1D DBM \citep{vrsnak13}, five versions of the drag-based model versions have been developed by the Hvar Observatory solar and heliospheric group in close collaboration with the solar and heliospheric group at the University of Graz. These five versions include three different geometries, as discussed in Section \ref{geometry}, and two different ensemble versions, as discussed in Section \ref{DBEM}. We give an overview of these five DBM versions and their respective tools in Section \ref{overview} and demonstrate their application on a real event in Section \ref{example}.

\section{Overview of Drag-based model tools}
\label{overview}
\subsection{The basic description of the model}
\label{model}
The Drag-Based Model (DBM) tools are all based on the equation of motion analogous to the aerodynamic drag:
\begin{equation}
a(t)=-\gamma(v(t)-w)|v(t)-w|
\,,
\label{eq1}
\end{equation}
where $a(t)=\mathrm{d}^2 R(t)/\mathrm{d}t^2$ is the CME acceleration, $v(t)=\mathrm{d}R(t)/\mathrm{d}t$ is the CME speed, $R(t)$ is the heliospheric distance, $\gamma$ is the drag parameter which describes CME speed change-rate and is assumed to be constant, and $w$ is the solar wind speed, also assumed to be constant. Along with the initial properties of the CME, which can be obtained from the coronagraphic observation, $\gamma$ and $w$ have to be specified to obtain analytical solutions of Equation \ref{eq1} for a specific CME, $R(t)$ and $v(t)$, given by  \citet{vrsnak13}:
\begin{equation}
\begin{split}
R(t)=\frac{S}{\gamma}ln[1+S\gamma(v_0-w)t]+wt+R_0\\
v(t)=\frac{v_0-w}{1+S\gamma(v_0-w)t}+w
\,,
\end{split}
\label{eq2}
\end{equation}
where $v_0=v(t=0)$ is the initial CME speed, $R_0=R(t=0)$ the corresponding starting radial distance, and S is a sign function ($S=1$ for $v_0 > w$, $S=-1$ for $v_0 < w$). These solutions describe the time-dependent part of the drag-based CME propagation and are thus the same in all tools, regardless of different geometries.

We note that generally speaking $\gamma$ and $w$ are not constant in time. However, it can be shown that at a sufficient distance from the Sun $\gamma$ and $w$ become approximetely constant and may be represented by their asymptotic values, which are approximately equal to the values at 1 AU \citep[see][for details]{vrsnak07,vrsnak13,zic15,manchester17}. Theoretically, the distance at which $\gamma=const.$ and $w=const.$ assumptions should hold is beyond $\approx15\mathrm{R_{\odot}}$ \citep{zic15}. On the other hand, the distance at which the drag force becomes dominant varies case-to-case, and is farther away for slower CMEs \citep{vrsnak01,vrsnak04a,sachdeva15}. Therefore, $R_0\le15\mathrm{R_{\odot}}$ might not be an optimal choice for the model. The Lorentz force was found to generally peak between 1.65 and 2.45 $\mathrm{R_{\odot}}$ and becomes negligible for faster CMEs at $3.5-4\mathrm{R_{\odot}}$, for slower at $12-50\mathrm{R_{\odot}}$ \citep{sachdeva17}. As the optimal value for the starting radial distance the DBM tools recommendation is $\ge20\mathrm{R_{\odot}}$, as this assumption was shown to be valid for a number of cases \citep{vrsnak10,vrsnak13} and should hold unless the CME is very slow and/or has a prolonged acceleration phase \citep{vrsnak01,vrsnak04a,sachdeva15,sachdeva17}. Nevertheless, it is recommended to always check whether the early CME kinematics indicates that the starting radial distance $\ge20\mathrm{R_{\odot}}$ is suitable for the observed CME.

\hspace{0.5cm}
\subsubsection{Using empirical $w$ and $\gamma$ values in DBM}
\label{empirical}
An important DBM issue is how to determine the input for $w$ and $\gamma$. There are several options to determine the input for $w$: 1) using an empirical value determined from statistical analysis; 2) using a solar wind model (e.g. numerical heliospheric model, an empirical model based on CH observation or persistence model); 3) using the solar-wind speed based on the in-situ measurements at 1 AU at the time of the ICME take-off. Using $w$ based on the in-situ measurements at the time of the ICME take-off was shown to be the same or even worse than using empirically obtained values \citep{vrsnak13}. Statistical analysis has shown that the most appropriate values for $w$ should be in the range $300-600\,\mathrm{km\,s}^{-1}$, with $w=500\,\mathrm{km\,s}^{-1}$ as the optimal value \citep[i.e. applicable to the broadest subset of CMEs][]{vrsnak13}. However, the optimal empirically derived value is sample-dependent and was found to be lower for a different sample \citep{vrsnak14}. Therefore, as an optimal empirically-based value $w=450\,\mathrm{km\,s}^{-1}$ is set for all tools. Recent analysis has shown that this value seems optimal even during the conditions of low solar activity \citep{calogovic20}. It should be noted that this value might not be valid if there is an equatorial coronal hole in the vicinity of the CME source region, where one should apply a higher value to take into account CME propagation through the high speed stream. For that purpose one can use a model of the solar wind speed, where empirical solar wind models are especially suitable, due to their simplicity and speed. DBM tools available at the European Space Agency (ESA) Space Situational Awareness (SSA) portal can be coupled with the Empirical Solar Wind Forecast tool, which is based on empirical modelling of the high-speed stream (HSS) arrival derived from coronal hole area observations \citep[see][]{vrsnak07a,temmer07,rotter12,reiss16}.

The $\gamma$ parameter is given by the expression \citep[e.g.][]{vrsnak13}:
\begin{equation}
\gamma=\frac{c_\mathrm{d}A\rho_\mathrm{w}}{M+M_\mathrm{v}}=\frac{c_\mathrm{d}}{L(\frac{\rho}{\rho_\mathrm{w}}+\frac{1}{2})}
\,,
\label{eq3}
\end{equation}
where $A$ is the CME cross-sectional area, $\rho_\mathrm{w}$ is the solar-wind density, $M$ is the CME mass, $M_\mathrm{v}$ is the so-called virtual mass (i.e. the mass of the material piled-up in front of the CME, $L$ is the CME thickness in the radial direction, $\rho$ is the CME density and $c_\mathrm{d}$ is the dimensionless drag coefficient, which in the DBM tools is taken to be 1 according to \citep{cargill04}. Theoretically, it is possible to estimate relative CME mass density and radial size to determine $\gamma$ based on coronagraphic measurements. However, the errors corresponding to these estimations \citep[$\approx 15\%$ for the mass][]{bein13} can yield $\gamma$ with a very large uncertainty. For a CME several times denser than the surrounding corona (e.g. $\rho/\rho_{\mathrm{w}}\approx5$), and of radial size $1-10\mathrm{R}_{\odot}$ one finds approximate range of $\gamma=0.2-2\cdot10^{-7}\,\,\mathrm{km}^{-1}$, which roughly corresponds to the range obtained from statistical analysis \citep{vrsnak13}. The distribution of the $\gamma$ obtained from statistical analysis is highly asymmetrical and weighted towards the lower values \citep{vrsnak13}, where $\gamma=0.2\cdot10^{-7}\,\,\mathrm{km}^{-1}$ was found as an optimal value in combination with $w=450\,\mathrm{km\,s}^{-1}$ \citep{vrsnak13,vrsnak14}. Therefore, this value has been chosen as optimal empirically-based value for DBM tools (of course customized values are allowed).

In addition, some of the DBM tools offer $\gamma$ options for slower and faster CMEs. Observationally, CME peak speed is related to the peak soft X-ray flux \citep{vrsnak05,maricic07}, and the flare fluence to the CME mass \citep{yashiro09,dissauer18}. This is interpreted in the context of a  feedback relationship between the CME dynamics and the reconnection process in the wake of the CME \citep{vrsnak16}. Consequently, we would expect faster CMEs to be more massive and thus expect lower $\gamma$ for faster CMEs and higher $\gamma$ for slower CMEs. Additional empirical-based fine-tuning of the $\gamma$ parameter may be performed by the user according to the relative CME brightness in the coronagraphic images, which is generally related to the CME mass \citep[see e.g.][and references therein]{colaninno09}. Massive CMEs are generally observed as brighter objects in the coronagraphic images, therefore one may use a lower value of  $\gamma$ in case of very bright CMEs or increase it for very faint CMEs. However, one needs to keep in mind that the observed intensity of a CME (and thus mass calculation) depends on the angle between the line-of-sight of the observer and the plane-of-sight, i.e. CME direction with respect to the Thomson surface \citep[for details see][]{howard09,colaninno09}. Finally, fine-tuning of the $\gamma$ parameter may be performed to account for the pre-conditioning of the interplanetary space due to preceding CME(s). Namely, preceding CME(s) may "deplete`` the heliospheric sector before the CME in question, resulting in lower density and thus lower drag forces \citep{temmer15,temmer17,desai20}. This effect can be taken into account by using a lower value for the $\gamma$ parameter \citep[see e.g.][]{temmer15,dumbovic19}. However, when "customising`` $\gamma$ one needs to be careful not to underestimate or overestimate it, as this can lead to underestimation or overestimation of the transit time, respectively. It was recently shown by \citet{paouris21} that underestimated $\gamma$ can lead to significant underestimation of the transit time, even if $w$ is underestimated, especially for the fast CMEs.

\hspace{0.5cm}
\subsubsection{Running DBM for shock propagation}
\label{shock}

While propagating in the interplanetary space CMEs may or may not drive shocks, however if they do, the arrival of the CME magnetic structure (i.e. ejected twisted magnetic structure) is preceded by the shock arrival \citep[for ICME overview see e.g.][]{zurbuchen06,kilpua17}. Physically, the DBM equation of motion describes the propagation of the CME magnetic structure and not of the associated shock. However, the comparison of the DBM with the heliospheric model ENLIL \citep{odstrcil04}, in which the CME is initiated as a pressure pulse and thus more suitable to track the shock front, has shown that there is in general a good agreement between the two when a lower value of the $\gamma$ parameter is applied \citep{vrsnak14}. Moreover, \citet{hess15} have found that both the shock front and the CME leading edge can be modelled in the heliosphere with a drag model, where the CME ejecta front undergoes a more rapid deceleration than the shock front and the propagation of the two fronts is not completely coupled in the heliosphere. Indeed, some drag-based models such as the ElEvo \citep{mostl15} and ElEvoHi \citep{rollett16} standardly follow the shock front. \citet{dumbovic18a} have also used a lower $\gamma$ value ($\gamma=0.1\cdot10^{-7}\,\mathrm\,{km}^{-1}$) to apply the DBM ensemble version to simulate CME shock propagation, whereas \citet{temmer15} and \citet{guo18b} have used DBM to model both shock and CME propagation separately, using both different input and a lower $\gamma$ value for the shock propagation. Therefore, we note that DBM tools can be used to simulate both CME and shock propagation, however it is important to keep in mind that: 1) the shock propagation is not necessarily coupled to CME propagation; 2) proper CME/shock input is used and 3) lower $\gamma$ values should be applied to the shock as compared to the CME propagation.

\hspace{0.5cm}
\subsection{DBM tools with different geometries}
\label{geometry}

\begin{figure}
\centerline{\includegraphics[width=1\textwidth]{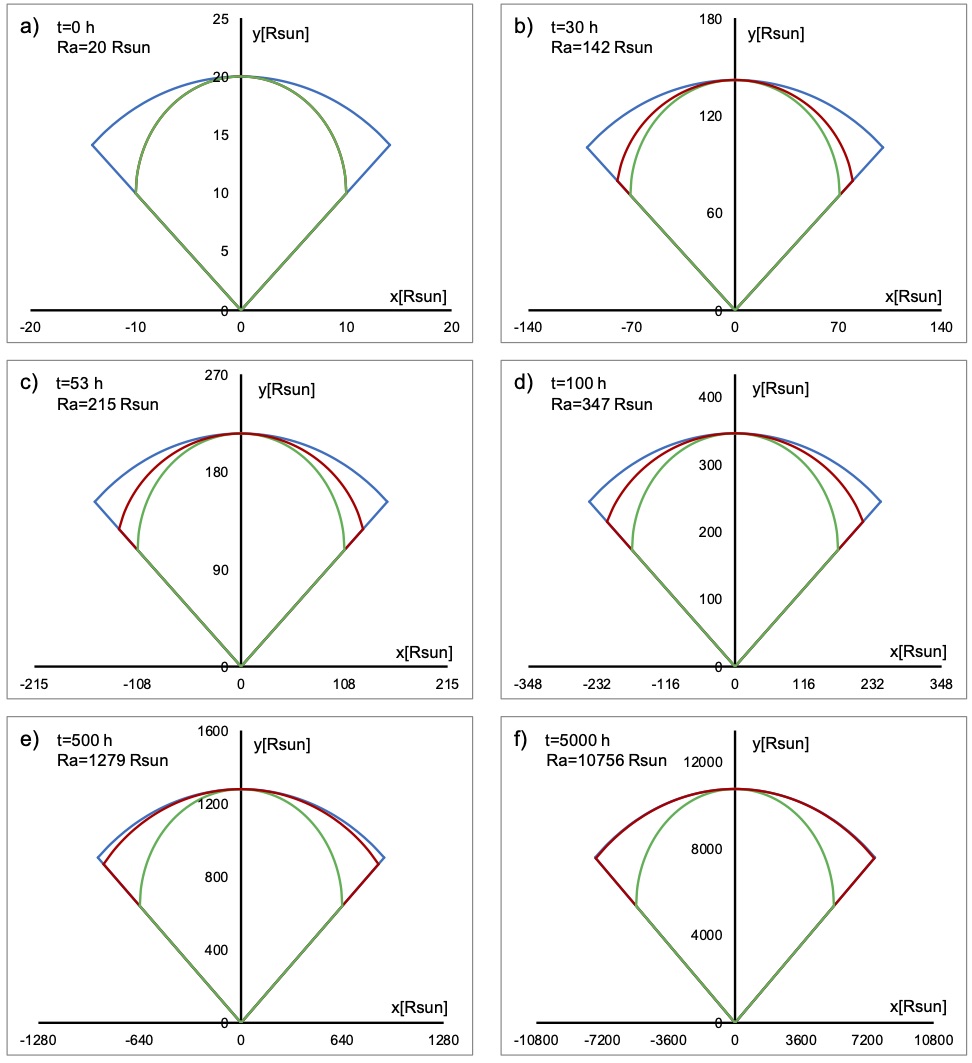}}
\caption{The differences between the fronts at 6 arbitrarily chosen time-steps for the three DBM tools: basic 1D DBM (the concentric leading edge, blue), advanced 2D DBM with self-similar cone geometry (green) and the advanced 2D DBM with flattening cone geometry (red). The subplots show the position of the leading edge in the XY coordinate system (i.e. the solar equatorial plane). The heliospheric distance of the apex, $R_a$ is highlighted in each time-step. The following DBM parameters were used to create the plots: initial CME speed of $1000 \mathrm\,{km\,s}^{-1}$, initial distance of $20 \mathrm{R}_{\odot}$, $\gamma$ of $0.2\cdot10^{-7} \mathrm\,{km}^{-1}$ and solar wind speed $w$ of $450 \mathrm\,{km\,s}^{-1}$.}
\label{fig1}
\end{figure}

\begin{figure}
\centerline{\includegraphics[width=1\textwidth]{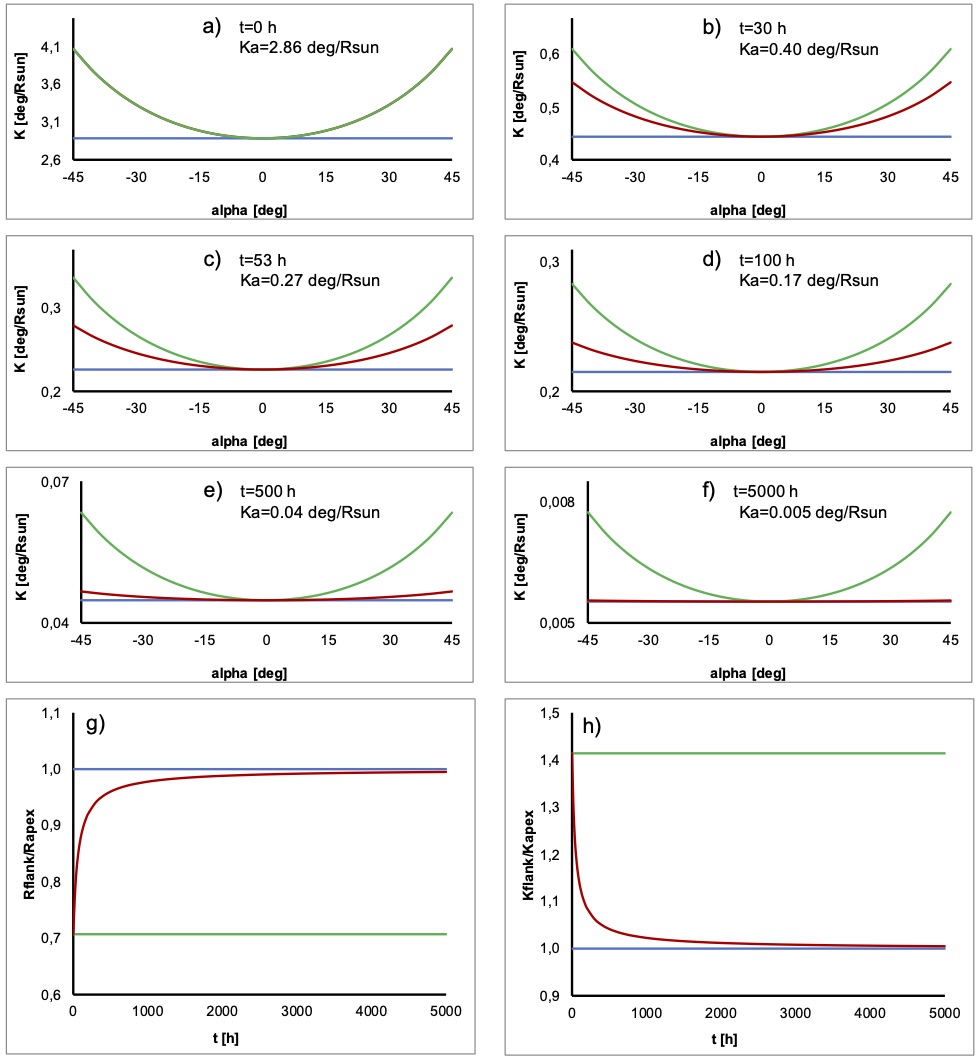}}
\caption{The differences between the curvatures of the fronts at different time-steps for the three DBM tools: basic 1D DBM (the concentric leading edge, blue), advanced 2D DBM with self-similar cone geometry (green) and the advanced 2D DBM with flattening cone geometry (red). The time-steps and DBM parameters correspond to those used in Figure \ref{fig1}. The subplots (a-f) show how the curvature $K$ along the leading edge behaves in time. The curvature at the apex, $K_a$ is highlighted in each time-step. The subplot g) shows how the ratio of the flank heliospheric distance and apex heliospheric distance ($R_f/R_a$) evolves in time, whereas subplot h) shows how the ratio of the curvature at the flank and at the apex ($K_f/K_a$) evolves in time.}
\label{fig2}
\end{figure}

The basic form of the DBM was formulated by \citet{vrsnak07,vrsnak10} and analysed in detail by \citep{vrsnak13}, where the basic 1D DBM tool was first presented. The basic 1D version of DBM is available as an online tool at Hvar Observatory webpage\footnote{\url{http://oh.geof.unizg.hr/DBM/dbm.php}} and relies on solutions given in Equation \ref{eq2}. Since it is a 1D equation it considers propagation of a single point, i.e. CME apex. The tool is also applicable to determine the propagation of an arbitrary, non-apex point of the CME leading edge, assuming that the CME leading edge evolves self-similarly as a circular arc concentric with the solar surface (i.e. all elements of the ICME front have the same heliocentric distance).  As can be seen in Figures \ref{fig1} and \ref{fig2}, this concentric geometry results in a self-similarly evolving CME leading edge. However, since the tool does not consider CME angular extent or its direction, it does not provide information whether or not this point hits a specific target. The basic assumptions, input, output and tool specifications are given in the second column of Table \ref{tab1}.

The advanced form of the DBM was formulated by \citep{zic15}, who applied a 2D cone geometry to the basic 1D DBM solutions given in Equation \ref{eq2}. The cone geometry was selected as it is a standard geometry used in heliospheric models such as ENLIL \citep{odstrcil04} or EUHFORIA \citep{pomoell18} and therefore their input would be suitable for use in DBM as well. The cone angular dependence is introduced in DBM in the following form:
\begin{equation}
\begin{split}
R(\alpha)=R_0\frac{\cos\alpha+\sqrt{\tan^2\omega-\sin^2\alpha}}{1+\tan\omega}\\
v(\alpha)=v_0\frac{\cos\alpha+\sqrt{\tan^2\omega-\sin^2\alpha}}{1+\tan\omega}
\,,
\end{split}
\label{eq4}
\end{equation}
where $R_0$ and $v_0$ are distance and speed of the plasma element at the CME apex, $\omega$ is the half-width of the cone (i.e. of the CME opening angle) and $\alpha$ is the opening angle corresponding to the plasma element in question. Depending on the application of the cone-geometry given by Equation \ref{eq4} to the basic 1D DBM solutions given in Equation \ref{eq2}, two different evolutions of the CME leading edge are possible, self-similar cone evolution and the flattening cone evolution. 

The self-similar evolution of the cone leading edge is obtained assuming that the CME front does not change its shape, i.e. when the DBM solutions for a plasma element after time $t$ at the angular distance $\alpha$ from the apex at the leading edge is given by:
\begin{equation}
\begin{split}
R(\alpha, t)=R_0(t)\frac{\cos\alpha+\sqrt{\tan^2\omega-\sin^2\alpha}}{1+\tan\omega}\\
v(\alpha, t)=v_0(t)\frac{\cos\alpha+\sqrt{\tan^2\omega-\sin^2\alpha}}{1+\tan\omega}
\,,
\end{split}
\label{eq5}
\end{equation}
where $R_0(t)$ and $v_0(t)$ are given by Equation \ref{eq2}. The self-similar cone leading edge is compared to the concentric geometry as well as the flattening cone leading edge in Figures \ref{fig1} and \ref{fig2}. It has been adopted in the online DBM tool that runs on the Hvar Observatory webpage\footnote{\url{http://oh.geof.unizg.hr/DBM/dbm.php}}, as well as the Community Coordinated Modelling Centre (CCMC)\footnote{\url{https://ccmc.gsfc.nasa.gov}}. Since the tool does implement information on the CME angular extent and its direction, it also provides information of whether or not the CME hits the target.The basic assumptions, input, output and tool specifications are given in the third column of Table \ref{tab1}.

The flattening cone leading edge evolution is obtained by propagating each plasma element of the CME leading edge independently, using the CME 2D cone geometry given by Equation \ref{eq4} as the initial leading edge. The DBM solutions for a plasma element after time $t$ at the angular distance $\alpha$ from the apex at the leading edge is given by:
\begin{equation}
\begin{split}
R(\alpha,t)=\frac{S}{\gamma}ln[1+S\gamma(v_0(\alpha)-w)t]+wt+R_0(\alpha)\\
v(\alpha,t)=\frac{v_0(\alpha)-w}{1+S\gamma(v_0(\alpha)-w)t}+w
\,,
\end{split}
\label{eq6}
\end{equation}
where $R_0(\alpha)$ and $v_0(\alpha)$ are given by Equation \ref{eq5}. The flattening cone leading edge is also shown in Figures \ref{fig1} and \ref{fig2} and similarly as 2D self-similar DBM provides information of whether or not the CME hits the target. The basic assumptions, input, output and tool specifications are given in the fourth column of Table \ref{tab1}.

To summarize, three different geometries of the CME leading edge (CME front) are considered in DBM tools: concentric arc, self-similarly evolving cone, and flattening cone. The differences between the fronts and their evolution for the three tools described above are shown in Figure \ref{fig1} for halfwidth $<90^{\circ}$ at several arbitrarily chosen time-steps. The 6 subplots (a-f) show the position of the leading edge in the XY coordinate system for 6 different time-steps. It can be seen that initially (at $t=0$) we differ only 2 geometries, the concentric arc and cone geometry leading edge. Although the initial shape of the flattening cone leading edge is that of a 2D cone, at $t>0$ the shape of the leading edge starts to deviate from the initial cone shape more and more. This is because each plasma element of the leading edge is propagated independently using different initial parameters. A plasma element at the flank will have a lower value of the initial speed than e.g. a plasma element at the apex and will therefore experience less drag if the CME is faster than the solar wind and more drag if the CME is slower than the solar wind. Since the drag will not act equally on each plasma element across the leading edge, the evolution of the leading edge will not be self-similar. Instead, as can be seen in Figure \ref{fig1}, during the evolution the leading edge will gradually change from the initial cone shape towards a flatter shape. 

This can be seen more prominently in Figure \ref{fig2}, which shows the time-evolution of the curvature of the CME leading edge with respect to the center of the Sun, calculated as $K=\mathrm{\Delta}\Theta/\mathrm{\Delta}L$, where $\mathrm{\Delta}\Theta=|\Theta_1-\Theta_2|$ is the angular distance and $\mathrm{\Delta}L=\int_{\Theta_1}^{\Theta_2} \sqrt{r^2+(\mathrm{d}r/\mathrm{d}\Theta)^2} \,\mathrm{d}\Theta$ is the corresponding arc length of the curve in polar coordinates. Note that thus defined $K$ does not correspond to the standard mathematical term curvature, which is defined with respect to the center of the circle and thus remains always constant across the circular arc. Instead, we define quantity $K$ to differ between the concentric arc and self-similar cone in the polar coordinates with origin at the center of the Sun. We can see that $K$ of the concentric arc is constant across the leading edge, whereas $K$ of the 2D cone at the apex is identical to that of the concentric arc, but then increases towards the flanks. However, the difference in $K$ between flank and apex remains constant in time for a self-similarly evolving cone front, whereas it reduces for the flattening cone front. 

The last two subplots of Figure \ref{fig2} show the ratio of the flank distance to the apex distance, $R_f/R_a$ and the ratio of the curvature at the flank and at the apex, $K_f/K_a$. For self-similarly evolving fronts $R_f/R_a$ and $K_f/K_a$ are constant and in the specific case of a concentric leading edge both equal 1 (values at the flank are equal to the values at the apex). We see that the apex evolves identically in all three cases. For a self-similarly evolving cone $R_f/R_a$ and $K_f/K_a$ remain constant. For the flattening cone $R_f/R_a$ and $K_f/K_a$ are not constant, as the $R_f/R_a$ increases and $K_f/K_a$ decreases in time, both approaching the values for the concentric leading edge. It should be noted however, that they never actually reach the values for the concentric leading edge. This is because although the flank experiences different drag than the apex, it is slower than the apex. The difference between $R_f$ and $R_a$ is increasing, converging to a certain value, as the drag eventually adjusts the speed of both the apex and the flank to the ambient solar wind speed. As the distance from the Sun increases, the difference between $R_f$ and $R_a$ becomes very small compared to values of $R_f$ and $R_a$, therefore $R_f/R_a$ seems to converge to 1, although mathematically it will never reach it and the flattening cone will never truly become a concentric arc. It is also important to note that for halfwidth of $90^{\circ}$ all three fronts are semi-circles with the origin at the Sun center and thus evolve identically, as the concentric arc.

\hspace{0.5cm}
\subsection{The ensemble versions of the DBM}
\label{DBEM}
As noted in Section \ref{intro}, drag-based models are computationally efficient and thus widely used in probabilistic/ensemble modelling approaches. Ensemble forecasting takes into account the errors and uncertainties of the input to quantify the resulting uncertainties in the model predictions. The variability of an observational  input is introduced by making an ensemble, i.e. sets of CME observations to calculate a distribution of predictions and forecast the confidence in the likelihood of the prediction. This can be achieved in two ways: 1) by taking independently built sets of CME observations (e.g. as provided by different observers) or 2) by creating sets of CME observations (by e.g. using measurements and error estimations provided by a single observer). These two ensemble options were adopted in the Drag-based ensemble model (DBEM) \citep{dumbovic18a} and DBEMv3 web tool \citep{calogovic20}, respectively, which both use 2D DBM with flattening cone as a background physical model. The output of both tools is the probability of arrival, which is calculated as the ratio of the number of runs that predict a hit and total number of runs. Based on the runs that predict a hit, distributions of arrival time and speed are generated, where the calculated medians represent the likeliest arrival time and speed, and the uncertainty range is given by the $95\%$ confidence interval. We note that initially there was a DBEMv2, which was replaced by a more advanced DBEMv3.

For a single CME, DBEM uses an ensemble of $n$ measurements of the same CME, which may not be mutually related in any way (e.g. it might be obtained by different observers, different methods or even measurements from different instruments). Each ensemble member has the same weight. Moreover, there is no assumption that the CME measurements of a particular CME, i.e. ensemble member, are independent of each other or that their spread in values follow a certain distribution. The variability of solar wind speed, $w$, and drag parameter, $\gamma$, is taken into account by producing $m$ of their synthetic values. These synthetic values are combined with an ensemble of $n$ CME measurements, to give a final ensemble of $n\cdot m^2$ members as an input, which, after $n\cdot m^2$ runs, produces a distribution of $n\cdot m^2$ calculated CME transit times and arrival speeds. The synthetic values for $w$ and $\gamma$ are produced assuming that their real measurements follow a normal distribution with a mean value and standard deviation serving as the model input. A cumulative standard normal distribution is then generated, defined on an interval $[0,m-1]$, where $m$ is the number of synthetic measurements, also needed as the model input. The $m$ values which correspond to the integer values of the cumulative standard normal distribution are selected as synthetic measurements. This way, for identical distribution and $m$, the selection always results in an identical set of synthetic values, which include the tips of the distribution tail. Therefore, for small $m$ the distribution of chosen synthetic measurements is too heavily weighted to the tail compared to the normal distribution, and larger $m$ is needed for synthetic measurements to be weighted properly, $m>15$ \citep{dumbovic18a}. The basic assumptions, input, output and tool specifications of DBEM are given in the fifth column of Table \ref{tab1}.

In DBEMv3 the CME ensemble is not produced by the observer, but the tool. Observational input values and uncertainties are provided for CME input as well as $w$ and $\gamma$, from which the tool generates $m$ ensemble members. Each ensemble member is produced by randomly picking one value for each input parameter, assuming it follows a normal distribution with observational input value as mean and standard deviation derived from uncertainty (uncertainty$=3\sigma$). Due to this randomness (which cannot be controlled), the ensemble is not likely to be identical each time identical input is used, which produces small differences in the output of the model for different runs using identical input. However, for large ensembles, $m>10000$, the differences of the output are negligible \citep[see][]{calogovic20}. The basic assumptions, input, output and tool specifications of DBEMv3 are given in the last column of Table \ref{tab1}.

We note that in the DBEMv3 the CME input parameters are considered to be independent of each other and therefore the procedure is somewhat similar to error propagation. In DBEM the CME parameters within one measurement set are not necessarily independent of each other, \textit{CME sets} are independent of each other. This is important due to the nature of the model input used, i.e. obtaining CME input from coronagraphic measurements. Coronagraphs only display a projection of a 3D structure. Therefore, in order to derive parameters of a 3D CME, some assumptions need to be made on the CME geometry. These assumptions as well as their application can vary from observer-to-observer and result in CME measurement sets where the distribution of single parameter variability may differ substantially from the normal distribution. A single observer on the other hand, is more likely to provide CME measurements with errors that follow a normal distribution. While a single observer is more likely to bias the mean of the normal distribution of an input parameter and thus introduce errors, we note that in the near-real-time forecasting, where a quick estimation of the CME input is needed, DBEMv3 is more applicable, since it uses input provided by only one observer/method.

\section{Running the DBM tools: example event}
\label{example}

\begin{figure}
\centerline{\includegraphics[width=1\textwidth]{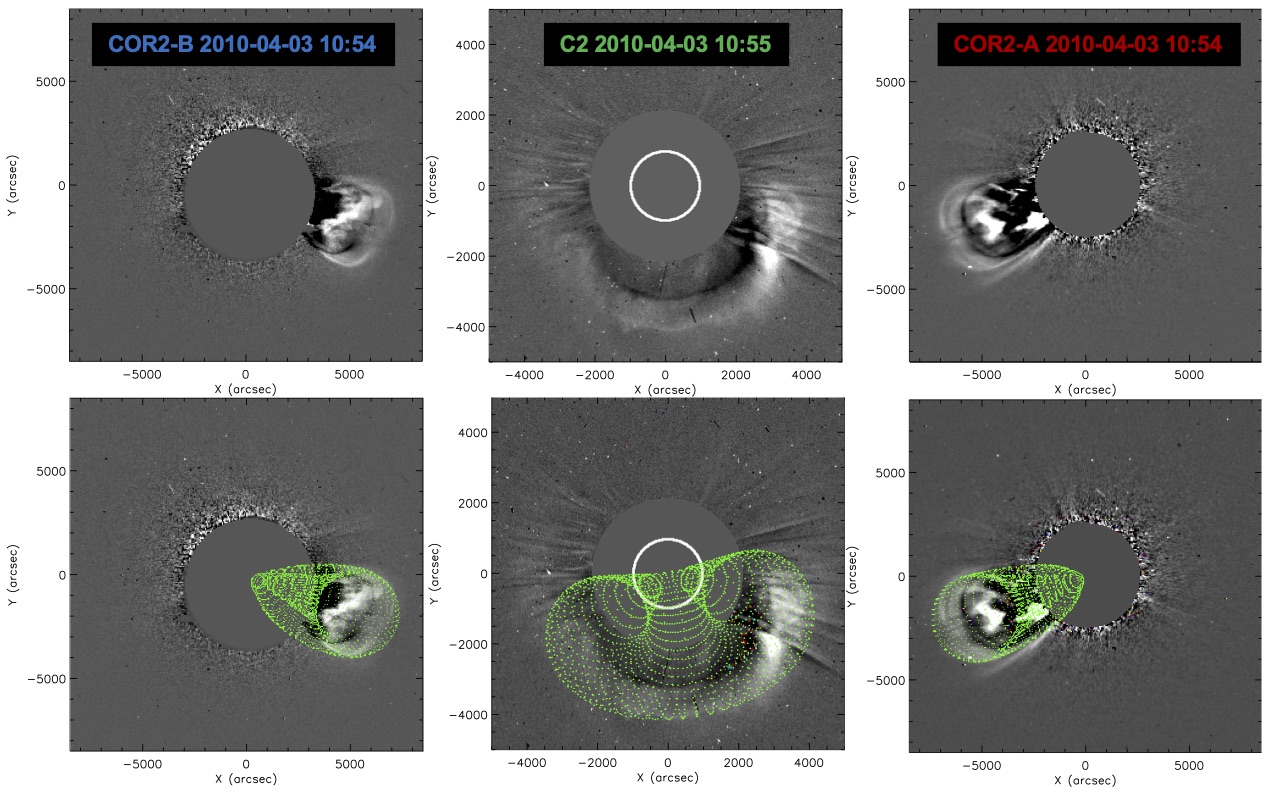}}
\caption{April 3rd 2010 CME observed in running difference images in STEREO/COR2 and SOHO/LASCO coronagraphs and the GCS reconstruction of the CME (green mesh).}
\label{fig3}
\end{figure}

We demonstrate the performance of DBM tools described in Section \ref{overview} by running all the tools using the same example event. As the example event we choose a previously studied CME that erupted on April 3rd 2010 and hit Earth on April 5th 2010 \citep[e.g.][]{mostl10,wood11,rodari18}. This event can be found in the SOHO/LASCO CME catalog \footnote{\url{https://cdaw.gsfc.nasa.gov/CME_list/}}, where it is listed as a halo with first appearance in LASCO-C2 April 3rd 2010 at 10:33 UT. In order to reconstruct the flux rope structure of the CME we use the Graduated cylindrical shell (GCS) model \citep{thernisien06,thernisien09,thernisien11}.

GCS is a geometrical model used to represent the flux rope structure of the CME to study its three-dimensional morphology, position, and kinematics. The flux rope is approximated with a self-similarly expanding hollow croissant originating from the center of the Sun, where the legs are conical, cross section circular, and the front pseudo-circular. The croissant is fully defined by six GCS parameters, which are: 1) longitude, 2) latitude, 3) height corresponding to the apex of the croissant, 4) the tilt of the croissant axis to the solar equatorial plane, 5) the croissant half-angle measured between the apex and the central axis of its leg, and 6) the ''aspect ratio`` (i.e. the sine of the angle defining the ''thickness`` of the croissant leg). The GCS parameters are obtained by fitting its 2D projections to the respective coronagraphic images, where at least 2 different vantage points are needed to constrain the geometry \citep[for further details on GCS see][]{thernisien06,thernisien09,thernisien11}.

To perform the reconstruction we use coronagraph images taken by SECCHI/COR2 \citep{howard08} onboard STEREO-A and B, as well as LASCO-C3 \citep{brueckner95} onboard SOHO spacecraft at April 3rd 2010 10:54 UT. Furthermore, the GCS reconstruction was done in 4 consecutive time-steps, following the CME leading edge while changing only the height parameter, i.e. assuming self-similar expansion. Based on these 4 measurements, a CME linear speed of $920\,\mathrm{km\,s}^{-1}$ was estimated. The GCS best fit parameters and linear speed are given in the second column of Table \ref{tab2} and the reconstruction is shown in Figure \ref{fig3}.

\begin{figure}
\centerline{\includegraphics[width=1\textwidth]{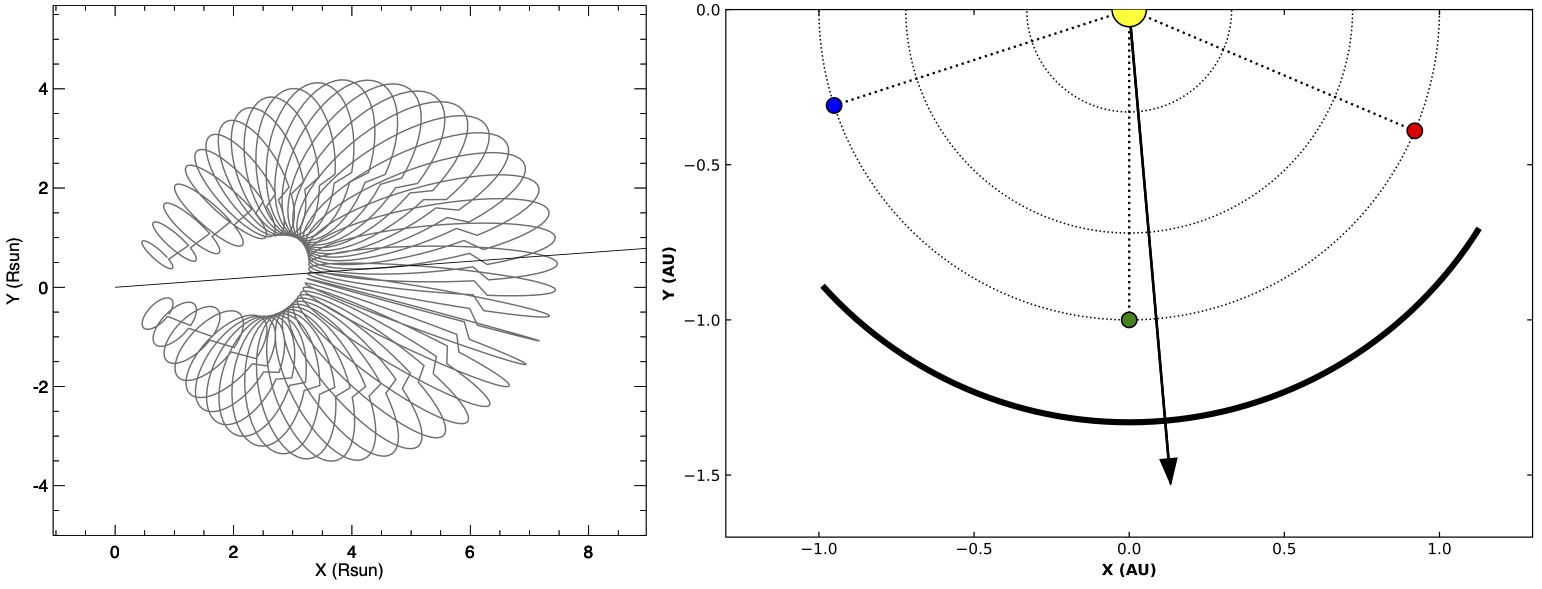}}
\caption{\textit{Left:} The projection of the GCS croissant of the April 3rd 2010 CME in the XY plane of the Heliocentric Earth Equatorial (HEEQ) coordinate system (i.e. plane of the solar equator). The direction of the apex is marked by a solid black line.
\textit{Right:} The direction and angular extent of the April 3rd 2010 CME (black arrow and black arc, respectively) and the positions of the three spacecraft in the HEEQ coordinate system at the CME liftoff time: STEREO-B (blue), SOHO (green), and STEREO-A (red).}
\label{fig4}
\end{figure}

Based on the results of the GCS reconstruction we derive input for the DBM tools (see column 4 in Table \ref{tab2}). Using the CME linear speed, we extrapolate the CME apex to $R_0=20\mathrm{R}_{\odot}$ assuming constant speed, which is taken as the initial CME speed, $v_0$. The CME angular extent (i.e. CME angular half width, $\lambda$) in the solar equatorial plane was estimated based on the GCS-derived tilt, as well as GCS face-on and edge-on widths, as described by \citet{dumbovic19} and adopted in DBEMv3. The projection of the GCS reconstructed CME in the solar equatorial plane is shown in Figure \ref{fig4}, as well as the calculated CME angular extent and the positions of the spacecraft. We can see that due to relatively small tilt and large half-angle, the angular extent of the CME is quite large. In addition, we can see that the direction of the apex (given by the longitude of the CME source region, $\phi_{\mathrm{CME}}$) is very close to the Sun-Earth line. We next run DBM tools for the input obtained from the GCS reconstruction (bottom rows of Table \ref{tab2}). In order to run DBEM, for which different sets of CME measurements are needed as input, we utilize measurements from previous studies on this event \citep{wood17,rodari18} and from online catalogs provided by the Advanced Forecast For Ensuring Communications Through Space (AFFECTS)\footnote{\url{http://www.affects-fp7.eu/home/}} catalogs. We use CME input provided by the AFFECTS-GCS database and AFFECTS-CAT database, where the latter is obtained with the CME Analysis Tool (CAT) modeling technique developed by \citet{millward13}. We also run all other DBM tools for these various CME inputs (as given in columns (5-8). The DBM input for these CME measurements was derived the same way as for the GCS reconstruction performed here. The CME speed for AFFECTS-GCS catalog was assumed to be the same as in other two GCS reconstructions. CME measurements provided by five different observers using three different measurement methods, as well as the corresponding five DBM inputs is given in Table \ref{tab2}.

The inner boundary, the drag parameter, solar wind speed and target distance are the same for each of the five DBM inputs and are $R_0=20\,\mathrm{R}_{\odot}$, $\gamma=0.2\cdot10^{-7}\,\,\mathrm{km}^{-1}$, $w=450\,\mathrm{km\,s}^{-1}$, and $R_{\mathrm{target}}=1$\,au, respectively. In DBEM, 15 synthetic values of $\gamma$ and $w$ were used within the uncertainty ranges $\pm0.1\cdot10^{-7}\,\mathrm{km}^{-1}$ and $\pm50\,\mathrm{km\,s}^{-1}$, respectively. In DBEMv3 default uncertainty ranges of the tool were used: $\pm30$\,min, $\pm0.1\cdot10^{-7}\,\mathrm{km}^{-1}$, $\pm50\,\mathrm{km\,s}^{-1}$, $\pm200\,\mathrm{km\,s}^{-1}$, $\pm15^\circ$, and $\pm30^\circ$ for lift-off time, $\gamma$, $w$, $v_0$, $\lambda$, and $\phi_{\mathrm{CME}}$, respectively, and 10000 runs were performed to obtain the results. It can be seen in Table \ref{tab2} that the difference in the output of different DBM tools is very similar for all DBM tools which use a single input set (basic 1D DBM, 2D self-similar DBM, 2D flattening cone DBM, and DBEMv3). This is because the direction of the apex is very close to the direction of the target, i.e. the CME is likely to hit the target close to the apex, where all geometries evolve similarly. This is also the reason for very high arrival probability, given that the most of the input sets consider a relatively wide CME. The input set derived based on measurements by \citet{wood17} is the only one yielding a DBEMv3 arrival probability $<100\%$, because the estimated CME half width is low compared to the uncertainty. The difference of the output is more prominent between different input sets than between different tools, with the exception of DBEM, which shows slightly different results compared to other tools. This is because, unlike other DBM tools, DBEM does not use a single input but takes into account the variability of different input sets.

The last two rows of Table \ref{tab2} show the observational results for the CME arrival, which are based on the CME-ICME association made by \citet{mostl10} and arrival time and speed values provided by the ICME catalog of \citet{richardson10}\footnote{\url{http://www.srl.caltech.edu/ACE/ASC/DATA/level3/icmetable2.htm}}, where the mean ICME speed is taken as arrival speed and ICME start (not the start of the disturbance) is taken as the arrival time. Measured ICME peak speed is also given for reference. We can see that all DBM outputs in Table \ref{tab2} overestimate arrival time by couple of hours and underestimate arrival speed by couple of tens of $\,\mathrm{km\,s}^{-1}$, indicated that the drag was overestimated. Indeed, as this is a fast event, 2D flattening cone DBM, DBEM and DBEMv3 would suggest the $\gamma=0.1\cdot10^{-7}\,\mathrm{km}^{-1}$ option. Running DBEMv3 and DBEM with $\gamma=0.1\cdot10^{-7}\,\mathrm{km}^{-1}$ and keeping other input identical as given in Table \ref{tab2} yields arrival times 2010-04-05 13:20 and 2010-04-05 11:50 UT, respectively, i.e. very close to the observed arrival time. The DBEMv3 and DBEM output for arrival speed is $705\,\mathrm{km\,s}^{-1}$ for both tools, which is closer to the observed ICME peak speed instead of the ICME mean speed. 

\section{Discussion and conclusion}
\label{conclusion}

The basic DBM equations (Equations \ref{eq1} and \ref{eq2}) describe CME propagation in a simple, physics-based and analytical way. Therefore, even when CME geometry is included (2D DBM) and in ensemble mode, the model runs very quickly (Table \ref{tab1}). With the development of different tools and their performance analysis, optimized DBM parameters (e.g. initial distance, solar wind speed and drag parameter) have been established, which are offered as default parameters in the DBM tools (Table \ref{tab1}). This makes the tools very easy to use even for the unexperienced users. On the other hand, the tools allow customized input for more experienced users (as described in Section \ref{empirical}). Therefore, DBM tools are simple to use and computationally efficient, which is their main advantage  compared to numerical MHD models.

DBM tools offer three different geometries, which identically describe the propagation of the CME apex, but differ in the description of the CME flanks, and might therefore differ in the applicability. For instance, both 2D DBM tools assume initial cone geometry, however one propagates it self-similarly, whereas the other does not. Therefore, it is reasonable to assume that 2D self-similar DBM might be more suitable for CMEs with (near) self-similar expansion. This might be the case with slower CMEs which show more symmetric in situ profiles \citep{masias-meza16}. Since they propagate with speeds closer to the solar wind speed, they experience less drag. On the other hand, 2D flattening cone DBM might be more suitable for very fast CMEs which show quite asymmetric in situ profiles \citep{masias-meza16}, indicating non-self similar expansion. They propagate with speeds much larger than the solar wind speed, and thus experience more drag.  The non-self similar vs. self-similar evolution might become even more important for considerations of 3D geometries. It is important to note that 2D flattening cone DBM does not consider change of the front in a manner that the flanks catch-up or overtake the apex. It is a purely geometrical effect, a change from a highly-curved cone geometry towards a less-curved concentric-arc-like geometry, and not related to e.g. non-homogeneous drag or internal forces that might cause the 'pancaking effect' \citep[e.g.][]{cargill94}. The cone geometry is also suitable to describe the propagation of a CME driven shock, which is typically faster and stronger at the nose compared to flanks \citep[e.g.][]{neugebauer13} and thus the flanks are 'delayed' with respect to the nose. On the other hand, for a freely propagating shock, assuming it propagates in homogeneous medium, a concentric arc geometry might be more suitable. As demonstrated on an example event, for CME propagation near the apex, all geometries and therefore all DBM tools show similar results (provided that the CME input is the same, see Table \ref{tab2}).

The ensemble options of the DBM provide a more comprehensive prediction compared to the other three tools, as they additionally calculate arrival probability and confidence interval of the arrival time and speed. In addition, although they rely on a large number of DBM runs ($>1000$), they are still computationally inexpensive  (Table \ref{tab1}). Therefore, they are quite useful from the aspect of space weather forecast and its evaluation. Since there is a difference in implementation of the CME input in DBEM and DBEMv3, their applicability may also differ. DBEMv3 tool is much faster and only needs input from one observer, thus it is easy to use in near-real time forecasts. On the other hand, DBEM may use various CME input sets, provided by different observers, different methods or even different instruments, and may thus be more suitable for evaluation purposes.

To summarize, this paper provides an overview of the assumptions, application and performance of the five DBM tools developed at Hvar Observatory. It is important to note that although these tools were developed sequentially and therefore each more recent tool contains improvements compared to the older version, the older versions still have their applicability.




\section*{Acknowledgments}
M.D. acknowledges support by the Croatian Science Foundation under the project IP-2020-02-9893 (ICOHOSS) and International Space Science Institute (ISSI) team ``Understanding Our Capabilities in Observing and Modeling Coronal Mass Ejections", led by C. Verbeke and M. Mierla. J.C., B. V. and D.S. acknowledge support by the Croatian Science Foundation under the project 7549 (MSOC). K.M. acknowledges support by the Croatian Science Foundation in the scope of the Young Researchers Career Development Project Training New Doctoral Students. This research has received financial support from the European Union's Horizon 2020 research and innovation program under grant agreement No. 824135 (SOLARNET).


\renewcommand{\arraystretch}{2}
\begin{landscape}
\begin{table}
\fontsize{9}{9}\selectfont
\caption{Comparison of DBM tools}
\label{tab1}
\begin{tabular}{|ccc|c|c|c|c|c|}
\hline
&					&					&\multirow{2}{*}{1D DBM}		&\multicolumn{2}{c|}{2D DBM}			&\multirow{2}{*}{DBEM}		&\multirow{2}{*}{DBEMv3}	\\
\cline{5-6}
&					&					&						&self-similar cone	&flattening cone	&						&					\\
\hline												
\multirow{4}{*}{\rotatebox[origin=c]{90}{BASIC}}&\multicolumn{1}{c|}{\multirow{4}{*}{\rotatebox[origin=c]{90}{ASSUMPTIONS}}}	%
					&drag parameter		&$\gamma=const$			&$\gamma=const$	&$\gamma=const$	&$\gamma=const$			&$\gamma=const$\\
\cline{3-8}												
&\multicolumn{1}{c|}{}	&solar wind speed		&$w=const$				&$w=const$		&$w=const$		&$w=const$				&$w=const$\\
\cline{3-8}													
&\multicolumn{1}{c|}{}	&self-similarity			&Y						&Y				&N				&N						&N\\
\cline{3-8}													
&\multicolumn{1}{c|}{}	&2D geometry			&concentric circular arc 		&ice cream cone	&ice cream cone	&ice cream cone			&ice cream cone\\
\hline
\multicolumn{2}{|c|}{\multirow{9}{*}{\rotatebox[origin=c]{90}{INPUT\tablefootnote{not including basic CME input which is the same for all tools: CME take-off date \& time, CME initial speed}}}}%
					&optimal $R_0 [\mathrm{R_{\odot}}]$&$R_0=20$		&$R_0=20$		&$R_0=20$		&$R_0=20$				&$R_0=20$\\
\cline{3-8}													
\multicolumn{2}{|c|}{}		&CME width input		&N						&Y				&Y				&Y						&Y\\									
\multicolumn{2}{|c|}{}		&source position input	&N						&Y				&Y				&Y						&Y\\
\multicolumn{2}{|c|}{}		&GCS input 			&N						&N				&N				&N						&Y\\		
\cline{3-8}													
\multicolumn{2}{|c|}{}		&optimal $w [\,\mathrm{km\,s}^{-1}]$&$w=450$			&$w=450$		&$w=450$		&$w=450$				&$w=450$\\
\multicolumn{2}{|c|}{}		&modelled $w$ option\tablefootnote{using the ESWF tool, see Section\ref{empirical}}
										&N						&N				&Y				&N						&Y\\
\cline{3-8}													
\multicolumn{2}{|c|}{}		&\multicolumn{1}{c|}{\multirow{3}{*}{optimal $\gamma [10^{-7}\,\mathrm{km}^{-1}]$}}%
										&\multicolumn{1}{c|}{\multirow{3}{*}{0.2}}%
																&\multirow{3}{*}{0.2}	&0.1 (fast CME)	&0.1 (fast CME)			&0.1 (fast CME)\\
\multicolumn{2}{|c|}{}		&\multicolumn{1}{c|}{}	&\multicolumn{1}{c|}{}		&\multicolumn{1}{c|}{}&0.2 (normal CME)	&0.2 (normal CME)			&0.2 (normal CME)\\
\multicolumn{2}{|c|}{}		&\multicolumn{1}{c|}{}	&\multicolumn{1}{c|}{}		&\multicolumn{1}{c|}{}&0.5 (slow CME)	&0.5 (slow CME)			&0.5 (slow CME)\\	
\hline
\multicolumn{2}{|c|}{\multirow{3}{*}{\rotatebox[origin=c]{90}{OUTPUT}}}%
					&CME arrival time		&Y						&Y				&Y				&Y						&Y\\									
\multicolumn{2}{|c|}{}		&CME arrival speed		&Y						&Y				&Y				&Y						&Y\\
\multicolumn{2}{|c|}{}		&CME arrival probability 	&N						&N				&N				&Y						&Y\\		
\hline
\multirow{3}{*}{\rotatebox[origin=c]{90}{TOOL}}&\multicolumn{1}{c|}{\multirow{3}{*}{\rotatebox[origin=c]{90}{SPECS}}}%
					&typical runtime		&0.5 sec					&2 sec			&2 sec			&$<2$ min (average PC)		&6 sec\\
\cline{3-8}	
\multicolumn{2}{|c|}{}		&source				&OH\tablefootnote{\url{http://oh.geof.unizg.hr/DBM/dbm.php}}%
																&OH, CCMC\tablefootnote{\url{https://ccmc.gsfc.nasa.gov}}%
																				&ESA SSA\tablefootnote{\url{http://swe.ssa.esa.int}}%
																								&run-on-request\tablefootnote{runs are available upon request to \href{mailto:mdumbovic@geof.hr}{mdumbovic@geof.hr}}%
																														&ESA SSA\\
\cline{3-8}	
\multicolumn{2}{|c|}{}		&reference			&\citep{vrsnak13}			&\citep{zic15}		&\citep{zic15}		&\citep{dumbovic18a}		&\citep{calogovic20}	\\													
\hline
\end{tabular}
\end{table}
\end{landscape}

\renewcommand{\arraystretch}{2}
\begin{table}
\fontsize{8}{6}\selectfont
\caption{CME measurements and the corresponding DBM input and output for the April 3rd 2010 CME for 5 different observers using 3 different methods.}
\label{tab2}
\begin{tabular}{|c|c|c|c|c|c|c|c|}
\hline
\multicolumn{3}{|c|}{}											&	observer 1 (GCS)	&	Rodari+2018 (GCS)		&	Wood+2017	&	AFFECTS-GCS	&	AFFECTS-CAT\\
\hline														
\multirow{16}{*}{\rotatebox[origin=c]{90}{INPUT}}
			&		&	obs time							&	04/03 10:54		&	04/03 11:24			&	04/03 10:00	&	04/03 12:08		&	04/03 12:08\\
			&		&	height [$\mathrm{R}_{\odot}$]			&	8.1				&	15.6					&	7.3			&	13.6				&	21.5\\
			&CME	&	lon [deg]							&	5				&	3					&	3			&	4				&	0\\
			&measurements
					&	lat [deg]							&	-26				&	-28.5					&	-16			&	-26				&	24.6\\
			&		&	tilt [deg]							&	11				&	1.7					&	-80			&	-1				&	--\\
			&		&	kappa							&	0.35				&	0.3					&	0.21			&	0.42				&	--\\
			&		&	halfangle [deg]						&	37				&	24.3					&	60			&	16				&	--\\
\cline{2-8}														
			&	\multirow{4}{*}{DBM input}
					&	liftoff time							&	04/03 13:24		&	04/03 12:19			&	04/03 12:30	&	04/03 13:29		&	04/03 11:47\\
			&		&	v0 [$\,\mathrm{km\,s}^{-1}$]				&	920				&	920					&	960			&	920				&	812\\
			&		&	lon [deg]							&	5				&	3					&	3			&	4				&	0\\
			&		&	half width [deg]						&	53				&	41					&	19			&	41				&	30\\
\hline
\hline
\multicolumn{2}{|c|}{ICME}&	ToA*								&	04/05 12:00		&	04/05 12:00			&	04/05 12:00	&	04/05 12:00		&	04/05 12:00\\
\multicolumn{2}{|c|}{OBSERVATION}		
					&	SoA** [$\,\mathrm{km\,s}^{-1}$]			&	640 (790)			&	640 (790)				&	640 (790)		&	640 (790)			&	640 (790)\\
\hline
\hline
\multirow{22}{*}{\rotatebox[origin=c]{90}{OUTPUT}}
			& 		&	ToA								&	04/05  17:39		&	04/05 16:34			&	04/05  16:45	&	04/05  17:44		&	04/05  19:43\\
			& basic	&	ToA $|O-C|$\,[h]***					&	5.7				&	4.6					&	4.7			&	5.7				&	7.7\\	
			& 1D DBM&	SoA [$\,\mathrm{km\,s}^{-1}$]			&	620				&	620					&	620			&	620				&	597\\
			&		&	SoA $|O-C|$ [$\,\mathrm{km\,s}^{-1}$]	&	20 (170)			&	20 (170)				&	20 (170)		&	20 (170)			&	43 (193)\\
\cline{2-8}														
			&		&	ToA								&	04/05  17:51		&	04/05  16:41			&	04/05  17:01	&	04/05  17:56		&	04/05  19:43\\
			& 2D self-similar
					&	ToA $|O-C|$\,[h]					&	5.9				&	4.7					&	5.0			&	5.9				&	7.7\\
			& cone DBM
					&	SoA [$\,\mathrm{km\,s}^{-1}$]			&	618				&	619					&	617			&	618				&	597\\
			&		&	SoA $|O-C|$ [$\,\mathrm{km\,s}^{-1}$]	&	22 (172)			&	21 (171)				&	23 (173)		&	22 (172)			&	43 (193)\\
\cline{2-8}														
			&		&	ToA								&	04/05  17:47		&	04/05  16:36			&	04/05  15:37	&	04/05  17:46		&	04/05  19:37\\
			& 2D flattening
					&	ToA $|O-C|$\,[h]					&	5.8				&	4.6					&	3.6			&	5.8				&	7.6\\
			& cone DBM
					&	SoA [$\,\mathrm{km\,s}^{-1}$]			&	619				&	620					&	627			&	620				&	598\\
			& 		&	SoA $|O-C|$ [$\,\mathrm{km\,s}^{-1}$]	&	21 (171)			&	20 (170)				&	13 (163)		&	20 (170)			&	42 (192)\\
\cline{2-8}														
			&	\multirow{7}{*}{DBEMv3}	
					&	arrival probability [\%]				&	100\%			&	100\%				&	91.5\%		&	100\%			&	99.2\%\\
			&		&	ToA								&	04/05  18:06		&	04/05  17:14			&	04/05 18:04	&	04/05  18:28		&	04/05  20:51\\
			&		&	ToA CI [h]****						&	+5.6/-5.1			&	+5.1/-5.8				&	+5.6/-6.9		&	+5.3/-5.8			&	+5.9/-7.7\\
			&		&	ToA $|O-C|$\,[h]					&	6.1				&	5.2					&	6.1			&	6.5				&	8.9\\
			&		&	SoA [$\,\mathrm{km\,s}^{-1}$]			&	618				&	616					&	612			&	615				&	591\\
			&		&	SoA CI [$\,\mathrm{km\,s}^{-1}$]		&	+64/-52			&	+70/-63				&	+54/-65		&	+51/-68			&	+51/-59\\
			&		&	SoA $|O-C|$ [$\,\mathrm{km\,s}^{-1}$]	&	22 (172)			&	24 (174)				&	28 (178)		&	25 (175)			&	49 (199)\\
\cline{2-8}														
			&	\multirow{5}{*}{DBEM}	
					&	arrival probability [\%]				&	\multicolumn{5}{c|}{100\%}\\	
			&		&	ToA								&	\multicolumn{5}{c|}{04/05  16:44}\\
			&		&	ToA CI [h]							&	\multicolumn{5}{c|}{+5.8/-5.4}\\			
			&		&	ToA $|O-C|$\,[h]					&	\multicolumn{5}{c|}{4.7}\\
			&		&	SoA [$\,\mathrm{km\,s}^{-1}$]			&	\multicolumn{5}{c|}{617}\\
			&		&	SoA CI [$\,\mathrm{km\,s}^{-1}$]		&	\multicolumn{5}{c|}{+97/-65}\\
			&		&	SoA $|O-C|$ [$\,\mathrm{km\,s}^{-1}$]	&	\multicolumn{5}{c|}{23 (173)}\\
\hline	
\end{tabular}
\begin{tablenotes}
\tiny
	\item[*] \textit{*ToA=time of arrival}
	\item[**] \textit{**SoA=speed of arrival; observed mean ICME speed (peak speed is given in brackets)}
	\item[***] \textit{***$|O-C|$=absolute value of the difference between observed and calculated values}
	\item[****] \textit{****CI=confidence interval (95\%)}
\end{tablenotes}
\end{table}

\bibliographystyle{frontiersinHLTH&FPHY}
\bibliography{REFs}

\begin{thebibliography}{82}
\expandafter\ifx\csname natexlab\endcsname\relax\def\natexlab#1{#1}\fi
\expandafter\ifx\csname urlstyle\endcsname\relax
  \expandafter\ifx\csname doi\endcsname\relax
  \def\doi#1{doi:\discretionary{}{}{}#1}\fi \else
  \expandafter\ifx\csname doi\endcsname\relax
  \def\doi{doi:\discretionary{}{}{}\begingroup \urlstyle{rm}\Url}\fi \fi
\expandafter\ifx\csname selectlanguage\endcsname\relax
  \def\selectlanguage#1{}\fi

\bibitem[{{Zhang} et~al.(2003){Zhang}, {Dere}, {Howard}, and
  {Bothmer}}]{zhang03}
{Zhang} J, {Dere} KP, {Howard} RA, {Bothmer} V.
\newblock {Identification of Solar Sources of Major Geomagnetic Storms between
  1996 and 2000}.
\newblock {\em Astrophys.~J.\/} {\bf 582} (2003) 520--533.
\newblock \doi{10.1086/344611}.

\bibitem[{{Bothmer} and {Schwenn}(1998)}]{bothmer98}
{Bothmer} V, {Schwenn} R.
\newblock {The structure and origin of magnetic clouds in the solar wind}.
\newblock {\em Ann.~Geophys.\/} {\bf 16} (1998) 1--24.
\newblock \doi{10.1007/s00585-997-0001-x}.

\bibitem[{{D{\'e}moulin} et~al.(2008){D{\'e}moulin}, {Nakwacki}, {Dasso}, and
  {Mandrini}}]{demoulin08}
{D{\'e}moulin} P, {Nakwacki} MS, {Dasso} S, {Mandrini} CH.
\newblock {Expected in Situ Velocities from a Hierarchical Model for Expanding
  Interplanetary Coronal Mass Ejections}.
\newblock {\em Solar~Phys.\/} {\bf 250} (2008) 347--374.
\newblock \doi{10.1007/s11207-008-9221-9}.

\bibitem[{{Siscoe} and {Schwenn}(2006)}]{siscoe06}
{Siscoe} G, {Schwenn} R.
\newblock {CME Disturbance Forecasting}.
\newblock {\em Space~Sci.~Rev.\/} {\bf 123} (2006) 453--470.
\newblock \doi{10.1007/s11214-006-9024-y}.

\bibitem[{Zhao and Dryer(2014)}]{zhao14}
Zhao X, Dryer M.
\newblock Current status of cme/shock arrival time prediction.
\newblock {\em Space Weather\/} {\bf 12} (2014) 448--469.
\newblock \doi{10.1002/2014SW001060}.
\newblock 2014SW001060.

\bibitem[{{Vourlidas} et~al.(2019){Vourlidas}, {Patsourakos}, and
  {Savani}}]{vourlidas19}
{Vourlidas} A, {Patsourakos} S, {Savani} NP.
\newblock {Predicting the geoeffective properties of coronal mass ejections:
  current status, open issues and path forward}.
\newblock {\em Philosophical Transactions of the Royal Society of London Series
  A\/} {\bf 377} (2019) 20180096.
\newblock \doi{10.1098/rsta.2018.0096}.

\bibitem[{{Zhang}(2020)}]{zhang20}
{Zhang} Jea.
\newblock {Earth-affecting Solar Transients: A Review of Progress in Solar
  Cycle 24}.
\newblock {\em In prep.\/}  (2020).

\bibitem[{{Gopalswamy} et~al.(2001){Gopalswamy}, {Lara}, {Yashiro}, {Kaiser},
  and {Howard}}]{gopalswamy01}
{Gopalswamy} N, {Lara} A, {Yashiro} S, {Kaiser} ML, {Howard} RA.
\newblock {Predicting the 1-AU arrival times of coronal mass ejections}.
\newblock {\em J.~Geophys.~Res.\/} {\bf 106} (2001) 29207--29218.
\newblock \doi{10.1029/2001JA000177}.

\bibitem[{{Paouris} and {Mavromichalaki}(2017)}]{paouris17}
{Paouris} E, {Mavromichalaki} H.
\newblock {Effective Acceleration Model for the Arrival Time of Interplanetary
  Shocks driven by Coronal Mass Ejections}.
\newblock {\em Solar~Phys.\/} {\bf 292} (2017) 180.
\newblock \doi{10.1007/s11207-017-1212-2}.

\bibitem[{{Dryer} et~al.(2001){Dryer}, {Fry}, {Sun}, {Deehr}, {Smith},
  {Akasofu} et~al.}]{dryer01}
{Dryer} M, {Fry} CD, {Sun} W, {Deehr} C, {Smith} Z, {Akasofu} SI, et~al.
\newblock {Prediction in Real Time of the 2000 July 14 Heliospheric Shock Wave
  and its Companions During the `Bastille' Epoch$^{*}$}.
\newblock {\em Solar~Phys.\/} {\bf 204} (2001) 265--284.
\newblock \doi{10.1023/A:1014200719867}.

\bibitem[{{Zhao} et~al.(2016){Zhao}, {Liu}, {Inhester}, {Feng}, {Wiegelmann},
  and {Lu}}]{zhao16a}
{Zhao} X, {Liu} YD, {Inhester} B, {Feng} X, {Wiegelmann} T, {Lu} L.
\newblock {Comparison of CME/Shock Propagation Models with Heliospheric Imaging
  and In Situ Observations}.
\newblock {\em Astrophys.~J.\/} {\bf 830} (2016) 48.
\newblock \doi{10.3847/0004-637X/830/1/48}.

\bibitem[{{Takahashi} and {Shibata}(2017)}]{takahashi17}
{Takahashi} T, {Shibata} K.
\newblock {Sheath-accumulating Propagation of Interplanetary Coronal Mass
  Ejection}.
\newblock {\em Astrophys.~J.~Lett.\/} {\bf 837} (2017) L17.
\newblock \doi{10.3847/2041-8213/aa624c}.

\bibitem[{{Sudar} et~al.(2016){Sudar}, {Vr{\v s}nak}, and
  {Dumbovi{\'c}}}]{sudar16}
{Sudar} D, {Vr{\v s}nak} B, {Dumbovi{\'c}} M.
\newblock {Predicting coronal mass ejections transit times to Earth with neural
  network}.
\newblock {\em Mon.~Not.~R.~Astron.~Soc.\/} {\bf 456} (2016) 1542--1548.
\newblock \doi{10.1093/mnras/stv2782}.

\bibitem[{{Liu} et~al.(2018){Liu}, {Ye}, {Shen}, {Wang}, and
  {Erd{\'e}lyi}}]{liu18}
{Liu} J, {Ye} Y, {Shen} C, {Wang} Y, {Erd{\'e}lyi} R.
\newblock {A New Tool for CME Arrival Time Prediction using Machine Learning
  Algorithms: CAT-PUMA}.
\newblock {\em Astrophys.~J.\/} {\bf 855} (2018) 109.
\newblock \doi{10.3847/1538-4357/aaae69}.

\bibitem[{{Wu} et~al.(2011){Wu}, {Dryer}, {Wu}, {Wood}, {Fry}, {Liou}
  et~al.}]{wu11}
{Wu} CC, {Dryer} M, {Wu} ST, {Wood} BE, {Fry} CD, {Liou} K, et~al.
\newblock {Global three-dimensional simulation of the interplanetary evolution
  of the observed geoeffective coronal mass ejection during the epoch 1-4
  August 2010}.
\newblock {\em Journal of Geophysical Research (Space Physics)\/} {\bf 116}
  (2011) A12103.
\newblock \doi{10.1029/2011JA016947}.

\bibitem[{{Odstrcil} et~al.(2004){Odstrcil}, {Riley}, and {Zhao}}]{odstrcil04}
{Odstrcil} D, {Riley} P, {Zhao} XP.
\newblock {Numerical simulation of the 12 May 1997 interplanetary CME event}.
\newblock {\em J.~Geophys.~Res.\/} {\bf 109} (2004) A02116.
\newblock \doi{10.1029/2003JA010135}.

\bibitem[{{Pomoell} and {Poedts}(2018)}]{pomoell18}
{Pomoell} J, {Poedts} S.
\newblock {EUHFORIA: European heliospheric forecasting information asset}.
\newblock {\em Journal of Space Weather and Space Climate\/} {\bf 8} (2018)
  A35.
\newblock \doi{10.1051/swsc/2018020}.

\bibitem[{{Miki{\'c}} et~al.(1999){Miki{\'c}}, {Linker}, {Schnack}, {Lionello},
  and {Tarditi}}]{mikic99}
{Miki{\'c}} Z, {Linker} JA, {Schnack} DD, {Lionello} R, {Tarditi} A.
\newblock {Magnetohydrodynamic modeling of the global solar corona}.
\newblock {\em Physics of Plasmas\/} {\bf 6} (1999) 2217--2224.
\newblock \doi{10.1063/1.873474}.

\bibitem[{{van der Holst} et~al.(2014){van der Holst}, {Sokolov}, {Meng},
  {Jin}, {Manchester}, {T{\'o}th} et~al.}]{holst14}
{van der Holst} B, {Sokolov} IV, {Meng} X, {Jin} M, {Manchester} I W~B,
  {T{\'o}th} G, et~al.
\newblock {Alfv{\'e}n Wave Solar Model (AWSoM): Coronal Heating}.
\newblock {\em Astrophys.~J.\/} {\bf 782} (2014) 81.
\newblock \doi{10.1088/0004-637X/782/2/81}.

\bibitem[{{Riley} et~al.(2018){Riley}, {Mays}, {Andries}, {Amerstorfer},
  {Biesecker}, {Delouille} et~al.}]{riley18}
{Riley} P, {Mays} ML, {Andries} J, {Amerstorfer} T, {Biesecker} D, {Delouille}
  V, et~al.
\newblock {Forecasting the Arrival Time of Coronal Mass Ejections: Analysis of
  the CCMC CME Scoreboard}.
\newblock {\em Space~Weather\/} {\bf 16} (2018) 1245--1260.
\newblock \doi{10.1029/2018SW001962}.

\bibitem[{{Zhang} et~al.(2006){Zhang}, {Baumjohann}, {Delva}, {Auster},
  {Balogh}, {Russell} et~al.}]{zhang06}
{Zhang} TL, {Baumjohann} W, {Delva} M, {Auster} HU, {Balogh} A, {Russell} CT,
  et~al.
\newblock {Magnetic field investigation of the Venus plasma environment:
  Expected new results from Venus Express}.
\newblock {\em Planet.~Space.~Sci.\/} {\bf 54} (2006) 1336--1343.
\newblock \doi{10.1016/j.pss.2006.04.018}.

\bibitem[{{Temmer}(2016)}]{temmer16}
{Temmer} M.
\newblock {Kinematical properties of coronal mass ejections}.
\newblock {\em Astron.~Nachr.\/} {\bf 337} (2016) 1010.
\newblock \doi{10.1002/asna.201612425}.

\bibitem[{{Sheeley} et~al.(1999){Sheeley}, {Walters}, {Wang}, and
  {Howard}}]{sheeley99}
{Sheeley} NR, {Walters} JH, {Wang} YM, {Howard} RA.
\newblock {Continuous tracking of coronal outflows: Two kinds of coronal mass
  ejections}.
\newblock {\em J.~Geophys.~Res.\/} {\bf 104} (1999) 24739--24768.
\newblock \doi{10.1029/1999JA900308}.

\bibitem[{{Gopalswamy} et~al.(2000){Gopalswamy}, {Lara}, {Lepping}, {Kaiser},
  {Berdichevsky}, and {St.~Cyr}}]{gopalswamy00}
{Gopalswamy} N, {Lara} A, {Lepping} RP, {Kaiser} ML, {Berdichevsky} D, {St~Cyr}
  OC.
\newblock {Interplanetary acceleration of coronal mass ejections}.
\newblock {\em Geophys.~Res.~Lett.\/} {\bf 27} (2000) 145--148.
\newblock \doi{10.1029/1999GL003639}.

\bibitem[{{Sachdeva} et~al.(2015){Sachdeva}, {Subramanian}, {Colaninno}, and
  {Vourlidas}}]{sachdeva15}
{Sachdeva} N, {Subramanian} P, {Colaninno} R, {Vourlidas} A.
\newblock {CME Propagation: Where does Aerodynamic Drag 'Take Over'?}
\newblock {\em Astrophys.~J.\/} {\bf 809} (2015) 158.
\newblock \doi{10.1088/0004-637X/809/2/158}.

\bibitem[{{Cargill}(2004)}]{cargill04}
{Cargill} PJ.
\newblock {On the Aerodynamic Drag Force Acting on Interplanetary Coronal Mass
  Ejections}.
\newblock {\em Solar~Phys.\/} {\bf 221} (2004) 135--149.
\newblock \doi{10.1023/B:SOLA.0000033366.10725.a2}.

\bibitem[{{Vr{\v s}nak} and {{\v Z}ic}(2007)}]{vrsnak07}
{Vr{\v s}nak} B, {{\v Z}ic} T.
\newblock {Transit times of interplanetary coronal mass ejections and the solar
  wind speed}.
\newblock {\em Astron.~Astrophys.\/} {\bf 472} (2007) 937--943.
\newblock \doi{10.1051/0004-6361:20077499}.

\bibitem[{{Cargill} et~al.(1996){Cargill}, {Chen}, {Spicer}, and
  {Zalesak}}]{cargill96}
{Cargill} PJ, {Chen} J, {Spicer} DS, {Zalesak} ST.
\newblock {Magnetohydrodynamic simulations of the motion of magnetic flux tubes
  through a magnetized plasma}.
\newblock {\em J.~Geophys.~Res.\/} {\bf 101} (1996) 4855--4870.
\newblock \doi{10.1029/95JA03769}.

\bibitem[{{Vr{\v s}nak} et~al.(2013){Vr{\v s}nak}, {{\v Z}ic}, {Vrbanec},
  {Temmer}, {Rollett}, {M{\"o}stl} et~al.}]{vrsnak13}
{Vr{\v s}nak} B, {{\v Z}ic} T, {Vrbanec} D, {Temmer} M, {Rollett} T,
  {M{\"o}stl} C, et~al.
\newblock {Propagation of Interplanetary Coronal Mass Ejections: The Drag-Based
  Model}.
\newblock {\em Solar~Phys.\/} {\bf 285} (2013) 295--315.
\newblock \doi{10.1007/s11207-012-0035-4}.

\bibitem[{{Vr{\v s}nak} et~al.(2014){Vr{\v s}nak}, {Temmer}, {{\v Z}ic},
  {Taktakishvili}, {Dumbovi{\'c}}, {M{\"o}stl} et~al.}]{vrsnak14}
{Vr{\v s}nak} B, {Temmer} M, {{\v Z}ic} T, {Taktakishvili} A, {Dumbovi{\'c}} M,
  {M{\"o}stl} C, et~al.
\newblock {Heliospheric Propagation of Coronal Mass Ejections: Comparison of
  Numerical WSA-ENLIL+Cone Model and Analytical Drag-based Model}.
\newblock {\em Astrophys.~J.~Suppl.\/} {\bf 213} (2014) 21.
\newblock \doi{10.1088/0067-0049/213/2/21}.

\bibitem[{{Hess} and {Zhang}(2014)}]{hess14}
{Hess} P, {Zhang} J.
\newblock {Stereoscopic Study of the Kinematic Evolution of a Coronal Mass
  Ejection and Its Driven Shock from the Sun to the Earth and the Prediction of
  Their Arrival Times}.
\newblock {\em Astrophys.~J.\/} {\bf 792} (2014) 49.
\newblock \doi{10.1088/0004-637X/792/1/49}.

\bibitem[{{Hess} and {Zhang}(2015)}]{hess15}
{Hess} P, {Zhang} J.
\newblock {Predicting CME Ejecta and Sheath Front Arrival at L1 with a
  Data-constrained Physical Model}.
\newblock {\em Astrophys.~J.\/} {\bf 812} (2015) 144.
\newblock \doi{10.1088/0004-637X/812/2/144}.

\bibitem[{{{\v Z}ic} et~al.(2015){{\v Z}ic}, {Vr{\v s}nak}, and
  {Temmer}}]{zic15}
{{\v Z}ic} T, {Vr{\v s}nak} B, {Temmer} M.
\newblock {Heliospheric Propagation of Coronal Mass Ejections: Drag-based Model
  Fitting}.
\newblock {\em Astrophys.~J.~Suppl.\/} {\bf 218} (2015) 32.
\newblock \doi{10.1088/0067-0049/218/2/32}.

\bibitem[{{M{\"o}stl} et~al.(2015){M{\"o}stl}, {Rollett}, {Frahm}, {Liu},
  {Long}, {Colaninno} et~al.}]{mostl15}
{M{\"o}stl} C, {Rollett} T, {Frahm} RA, {Liu} YD, {Long} DM, {Colaninno} RC,
  et~al.
\newblock {Strong coronal channelling and interplanetary evolution of a solar
  storm up to Earth and Mars}.
\newblock {\em Nature Communications\/} {\bf 6} (2015) 7135.
\newblock \doi{10.1038/ncomms8135}.

\bibitem[{{Rollett} et~al.(2016){Rollett}, {M{\"o}stl}, {Isavnin}, {Davies},
  {Kubicka}, {Amerstorfer} et~al.}]{rollett16}
{Rollett} T, {M{\"o}stl} C, {Isavnin} A, {Davies} JA, {Kubicka} M,
  {Amerstorfer} UV, et~al.
\newblock {ElEvoHI: A Novel CME Prediction Tool for Heliospheric Imaging
  Combining an Elliptical Front with Drag-based Model Fitting}.
\newblock {\em Astrophys.~J.\/} {\bf 824} (2016) 131.
\newblock \doi{10.3847/0004-637X/824/2/131}.

\bibitem[{{Kay} and {Gopalswamy}(2018)}]{kay18}
{Kay} C, {Gopalswamy} N.
\newblock {The Effects of Uncertainty in Initial CME Input Parameters on
  Deflection, Rotation, B$_{z}$, and Arrival Time Predictions}.
\newblock {\em J.~Geophys.~Res.\/} {\bf 123} (2018) 7220--7240.
\newblock \doi{10.1029/2018JA025780}.

\bibitem[{{M{\"o}stl} et~al.(2018){M{\"o}stl}, {Amerstorfer}, {Palmerio},
  {Isavnin}, {Farrugia}, {Lowder} et~al.}]{mostl18}
{M{\"o}stl} C, {Amerstorfer} T, {Palmerio} E, {Isavnin} A, {Farrugia} CJ,
  {Lowder} C, et~al.
\newblock {Forward Modeling of Coronal Mass Ejection Flux Ropes in the Inner
  Heliosphere with 3DCORE}.
\newblock {\em Space~Weather\/} {\bf 16} (2018) 216--229.
\newblock \doi{10.1002/2017SW001735}.

\bibitem[{{Napoletano} et~al.(2018){Napoletano}, {Forte}, {Moro},
  {Pietropaolo}, {Giovannelli}, and {Berrilli}}]{napoletano18}
{Napoletano} G, {Forte} R, {Moro} DD, {Pietropaolo} E, {Giovannelli} L,
  {Berrilli} F.
\newblock {A probabilistic approach to the drag-based model}.
\newblock {\em J.~Space~Weather~Space~Clim.\/} {\bf 8} (2018) A11.
\newblock \doi{10.1051/swsc/2018003}.

\bibitem[{{Dumbovi{\'c}} et~al.(2018){Dumbovi{\'c}}, {{\v C}alogovi{\'c}},
  {Vr{\v s}nak}, {Temmer}, {Mays}, {Veronig} et~al.}]{dumbovic18a}
{Dumbovi{\'c}} M, {{\v C}alogovi{\'c}} J, {Vr{\v s}nak} B, {Temmer} M, {Mays}
  ML, {Veronig} A, et~al.
\newblock {The Drag-based Ensemble Model (DBEM) for Coronal Mass Ejection
  Propagation}.
\newblock {\em Astrophys.~J.\/} {\bf 854} (2018) 180.
\newblock \doi{10.3847/1538-4357/aaaa66}.

\bibitem[{{Amerstorfer} et~al.(2018){Amerstorfer}, {M{\"o}stl}, {Hess},
  {Temmer}, {Mays}, {Reiss} et~al.}]{amerstorfer18}
{Amerstorfer} T, {M{\"o}stl} C, {Hess} P, {Temmer} M, {Mays} ML, {Reiss} MA,
  et~al.
\newblock {Ensemble Prediction of a Halo Coronal Mass Ejection Using
  Heliospheric Imagers}.
\newblock {\em Space~Weather\/} {\bf 16} (2018) 784--801.
\newblock \doi{10.1029/2017SW001786}.

\bibitem[{{Kay} et~al.(2020){Kay}, {Mays}, and {Verbeke}}]{kay20}
{Kay} C, {Mays} ML, {Verbeke} C.
\newblock {Identifying Critical Input Parameters for Improving Drag-Based CME
  Arrival Time Predictions}.
\newblock {\em Space~Weather\/} {\bf 18} (2020) e02382.
\newblock \doi{10.1029/2019SW002382}.

\bibitem[{{Amerstorfer} et~al.(2020){Amerstorfer}, {Hinterreiter}, {Reiss},
  {M{\"o}stl}, {Davies}, {Bailey} et~al.}]{amerstorfer20}
{Amerstorfer} T, {Hinterreiter} J, {Reiss} MA, {M{\"o}stl} C, {Davies} JA,
  {Bailey} RL, et~al.
\newblock {CME arrival prediction using ensemble modeling based on heliospheric
  imaging observations}.
\newblock {\em submitted to Space Weather\/}  (2020) arXiv:2008.02576.

\bibitem[{{Manchester} et~al.(2017){Manchester}, {Kilpua}, {Liu}, {Lugaz},
  {Riley}, {T{\"o}r{\"o}k} et~al.}]{manchester17}
{Manchester} W, {Kilpua} EKJ, {Liu} YD, {Lugaz} N, {Riley} P, {T{\"o}r{\"o}k}
  T, et~al.
\newblock {The Physical Processes of CME/ICME Evolution}.
\newblock {\em Space~Sci.~Rev.\/} {\bf 212} (2017) 1159--1219.
\newblock \doi{10.1007/s11214-017-0394-0}.

\bibitem[{{Vr{\v s}nak}(2001)}]{vrsnak01}
{Vr{\v s}nak} B.
\newblock {Dynamics of solar coronal eruptions}.
\newblock {\em J.~Geophys.~Res.\/} {\bf 106} (2001) 25249--25260.
\newblock \doi{10.1029/2000JA004007}.

\bibitem[{{Vr{\v s}nak} et~al.(2004){Vr{\v s}nak}, {Ru{\v z}djak}, {Sudar}, and
  {Gopalswamy}}]{vrsnak04a}
{Vr{\v s}nak} B, {Ru{\v z}djak} D, {Sudar} D, {Gopalswamy} N.
\newblock {Kinematics of coronal mass ejections between 2 and 30 solar radii.
  What can be learned about forces governing the eruption?}
\newblock {\em Astron.~Astrophys.\/} {\bf 423} (2004) 717--728.
\newblock \doi{10.1051/0004-6361:20047169}.

\bibitem[{{Sachdeva} et~al.(2017){Sachdeva}, {Subramanian}, {Vourlidas}, and
  {Bothmer}}]{sachdeva17}
{Sachdeva} N, {Subramanian} P, {Vourlidas} A, {Bothmer} V.
\newblock {CME Dynamics Using STEREO and LASCO Observations: The Relative
  Importance of Lorentz Forces and Solar Wind Drag}.
\newblock {\em Solar~Phys.\/} {\bf 292} (2017) 118.
\newblock \doi{10.1007/s11207-017-1137-9}.

\bibitem[{{Vr{\v s}nak} et~al.(2010){Vr{\v s}nak}, {{\v Z}ic}, {Falkenberg},
  {M{\"o}stl}, {Vennerstrom}, and {Vrbanec}}]{vrsnak10}
{Vr{\v s}nak} B, {{\v Z}ic} T, {Falkenberg} TV, {M{\"o}stl} C, {Vennerstrom} S,
  {Vrbanec} D.
\newblock {The role of aerodynamic drag in propagation of interplanetary
  coronal mass ejections}.
\newblock {\em Astron.~Astrophys.\/} {\bf 512} (2010) A43.
\newblock \doi{10.1051/0004-6361/200913482}.

\bibitem[{{{\v C}alogovi{\'c}} et~al.(2021){{\v C}alogovi{\'c}},
  {Dumbovi{\'c}}, {Sudar}, {Vr{\v s}nak}, {Martini{\'c}}, {Temmer}
  et~al.}]{calogovic20}
{{\v C}alogovi{\'c}} J, {Dumbovi{\'c}} M, {Sudar} D, {Vr{\v s}nak} B,
  {Martini{\'c}} K, {Temmer} M, et~al.
\newblock {Probabilistic Drag-Based Ensemble Model (DBEM) evaluation for
  heliospheric propagation of CMEs}.
\newblock {\em Solar~Phys.\/} {\bf {submitted}} (2021).

\bibitem[{{Vr{\v s}nak} et~al.(2007){Vr{\v s}nak}, {Temmer}, and
  {Veronig}}]{vrsnak07a}
{Vr{\v s}nak} B, {Temmer} M, {Veronig} AM.
\newblock {Coronal Holes and Solar Wind High-Speed Streams: I. Forecasting the
  Solar Wind Parameters}.
\newblock {\em Solar~Phys.\/} {\bf 240} (2007) 315--330.
\newblock \doi{10.1007/s11207-007-0285-8}.

\bibitem[{{Temmer} et~al.(2007){Temmer}, {Vr{\v{s}}nak}, and
  {Veronig}}]{temmer07}
{Temmer} M, {Vr{\v{s}}nak} B, {Veronig} AM.
\newblock {Periodic Appearance of Coronal Holes and the Related Variation of
  Solar Wind Parameters}.
\newblock {\em Solar~Phys.\/} {\bf 241} (2007) 371--383.
\newblock \doi{10.1007/s11207-007-0336-1}.

\bibitem[{{Rotter} et~al.(2012){Rotter}, {Veronig}, {Temmer}, and
  {Vr{\v{s}}nak}}]{rotter12}
{Rotter} T, {Veronig} AM, {Temmer} M, {Vr{\v{s}}nak} B.
\newblock {Relation Between Coronal Hole Areas on the Sun and the Solar Wind
  Parameters at 1 AU}.
\newblock {\em Solar~Phys.\/} {\bf 281} (2012) 793--813.
\newblock \doi{10.1007/s11207-012-0101-y}.

\bibitem[{{Reiss} et~al.(2016){Reiss}, {Temmer}, {Veronig}, {Nikolic},
  {Vennerstrom}, {Sch{\"o}ngassner} et~al.}]{reiss16}
{Reiss} MA, {Temmer} M, {Veronig} AM, {Nikolic} L, {Vennerstrom} S,
  {Sch{\"o}ngassner} F, et~al.
\newblock {Verification of high-speed solar wind stream forecasts using
  operational solar wind models}.
\newblock {\em Space~Weather\/} {\bf 14} (2016) 495--510.
\newblock \doi{10.1002/2016SW001390}.

\bibitem[{{Bein} et~al.(2013){Bein}, {Temmer}, {Vourlidas}, {Veronig}, and
  {Utz}}]{bein13}
{Bein} BM, {Temmer} M, {Vourlidas} A, {Veronig} AM, {Utz} D.
\newblock {The Height Evolution of the ``True'' Coronal Mass Ejection Mass
  derived from STEREO COR1 and COR2 Observations}.
\newblock {\em Astrophys.~J.\/} {\bf 768} (2013) 31.
\newblock \doi{10.1088/0004-637X/768/1/31}.

\bibitem[{{Vr\v snak} et~al.(2005){Vr\v snak}, {Sudar}, and {Ru\v
  zdjak}}]{vrsnak05}
{Vr\v snak} B, {Sudar} D, {Ru\v zdjak} D.
\newblock {The CME-flare relationship: Are there really two types of CMEs?}
\newblock {\em Astron.~Astrophys.\/} {\bf 435} (2005) 1149--1157.
\newblock \doi{10.1051/0004-6361:20042166}.

\bibitem[{{Mari{\v c}i{\'c}} et~al.(2007){Mari{\v c}i{\'c}}, {Vr{\v s}nak},
  {Stanger}, {Veronig}, {Temmer}, and {Ro{\v s}a}}]{maricic07}
{Mari{\v c}i{\'c}} D, {Vr{\v s}nak} B, {Stanger} AL, {Veronig} AM, {Temmer} M,
  {Ro{\v s}a} D.
\newblock {Acceleration Phase of Coronal Mass Ejections: II. Synchronization of
  the Energy Release in the Associated Flare}.
\newblock {\em Solar~Phys.\/} {\bf 241} (2007) 99--112.
\newblock \doi{10.1007/s11207-007-0291-x}.

\bibitem[{{Yashiro} and {Gopalswamy}(2009)}]{yashiro09}
{Yashiro} S, {Gopalswamy} N.
\newblock {Statistical relationship between solar flares and coronal mass
  ejections}.
\newblock {Gopalswamy} N, {Webb} DF, editors, {\em Universal Heliophysical
  Processes\/} (2009), vol. 257, 233--243.
\newblock \doi{10.1017/S1743921309029342}.

\bibitem[{{Dissauer} et~al.(2019){Dissauer}, {Veronig}, {Temmer}, and
  {Podladchikova}}]{dissauer18}
{Dissauer} K, {Veronig} AM, {Temmer} M, {Podladchikova} T.
\newblock {Statistics of Coronal Dimmings Associated with Coronal Mass
  Ejections. II. Relationship between Coronal Dimmings and Their Associated
  CMEs}.
\newblock {\em Astrophys.~J.\/} {\bf 874} (2019) 123.
\newblock \doi{10.3847/1538-4357/ab0962}.

\bibitem[{{Vr{\v{s}}nak}(2016)}]{vrsnak16}
{Vr{\v{s}}nak} B.
\newblock {Solar eruptions: The CME-flare relationship}.
\newblock {\em Astronomische Nachrichten\/} {\bf 337} (2016) 1002.
\newblock \doi{10.1002/asna.201612424}.

\bibitem[{{Colaninno} and {Vourlidas}(2009)}]{colaninno09}
{Colaninno} RC, {Vourlidas} A.
\newblock {First Determination of the True Mass of Coronal Mass Ejections: A
  Novel Approach to Using the Two STEREO Viewpoints}.
\newblock {\em Astrophys.~J.\/} {\bf 698} (2009) 852--858.
\newblock \doi{10.1088/0004-637X/698/1/852}.

\bibitem[{{Howard} and {Tappin}(2009)}]{howard09}
{Howard} TA, {Tappin} SJ.
\newblock {Interplanetary Coronal Mass Ejections Observed in the Heliosphere:
  1. Review of Theory}.
\newblock {\em Space~Sci.~Rev.\/} {\bf 147} (2009) 31--54.
\newblock \doi{10.1007/s11214-009-9542-5}.

\bibitem[{{Temmer} and {Nitta}(2015)}]{temmer15}
{Temmer} M, {Nitta} NV.
\newblock {Interplanetary Propagation Behavior of the Fast Coronal Mass
  Ejection on 23 July 2012}.
\newblock {\em Solar~Phys.\/} {\bf 290} (2015) 919--932.
\newblock \doi{10.1007/s11207-014-0642-3}.

\bibitem[{{Temmer} et~al.(2017){Temmer}, {Reiss}, {Nikolic}, {Hofmeister}, and
  {Veronig}}]{temmer17}
{Temmer} M, {Reiss} MA, {Nikolic} L, {Hofmeister} SJ, {Veronig} AM.
\newblock {Preconditioning of Interplanetary Space Due to Transient CME
  Disturbances}.
\newblock {\em Astrophys.~J.\/} {\bf 835} (2017) 141.
\newblock \doi{10.3847/1538-4357/835/2/141}.

\bibitem[{{Desai} et~al.(2020){Desai}, {Zhang}, {Davies}, {Stawarz},
  {Mico-Gomez}, and {Iv{\'a}{\~n}ez-Ballesteros}}]{desai20}
{Desai} RT, {Zhang} H, {Davies} EE, {Stawarz} JE, {Mico-Gomez} J,
  {Iv{\'a}{\~n}ez-Ballesteros} P.
\newblock {Three-Dimensional Simulations of Solar Wind Preconditioning and the
  23 July 2012 Interplanetary Coronal Mass Ejection}.
\newblock {\em Solar~Phys.\/} {\bf 295} (2020) 130.
\newblock \doi{10.1007/s11207-020-01700-5}.

\bibitem[{{Dumbovi{\'c}} et~al.(2019){Dumbovi{\'c}}, {Guo}, {Temmer}, {Mays},
  {Veronig}, {Heinemann} et~al.}]{dumbovic19}
{Dumbovi{\'c}} M, {Guo} J, {Temmer} M, {Mays} ML, {Veronig} A, {Heinemann} SG,
  et~al.
\newblock {Unusual Plasma and Particle Signatures at Mars and STEREO-A Related
  to CME--CME Interaction}.
\newblock {\em Astrophys.~J.\/} {\bf 880} (2019) 18.
\newblock \doi{10.3847/1538-4357/ab27ca}.

\bibitem[{{Paouris} et~al.(2021){Paouris}, {{\v{C}}alogovi{\'c}},
  {Dumbovi{\'c}}, {Mays}, {Vourlidas}, {Papaioannou} et~al.}]{paouris21}
{Paouris} E, {{\v{C}}alogovi{\'c}} J, {Dumbovi{\'c}} M, {Mays} ML, {Vourlidas}
  A, {Papaioannou} A, et~al.
\newblock {Propagating Conditions and the Time of ICME Arrival: A Comparison of
  the Effective Acceleration Model with ENLIL and DBEM Models}.
\newblock {\em Solar~Phys.\/} {\bf 296} (2021) 12.
\newblock \doi{10.1007/s11207-020-01747-4}.

\bibitem[{{Zurbuchen} and {Richardson}(2006)}]{zurbuchen06}
{Zurbuchen} TH, {Richardson} IG.
\newblock {In-Situ Solar Wind and Magnetic Field Signatures of Interplanetary
  Coronal Mass Ejections}.
\newblock {\em Space~Sci.~Rev.\/} {\bf 123} (2006) 31--43.
\newblock \doi{10.1007/s11214-006-9010-4}.

\bibitem[{{Kilpua} et~al.(2017){Kilpua}, {Koskinen}, and
  {Pulkkinen}}]{kilpua17}
{Kilpua} E, {Koskinen} HEJ, {Pulkkinen} TI.
\newblock {Coronal mass ejections and their sheath regions in interplanetary
  space}.
\newblock {\em Living Reviews in Solar Physics\/} {\bf 14} (2017) 5.
\newblock \doi{10.1007/s41116-017-0009-6}.

\bibitem[{{Guo} et~al.(2018){Guo}, {Dumbovi{\'c}}, {Wimmer-Schweingruber},
  {Temmer}, {Lohf}, {Wang} et~al.}]{guo18b}
{Guo} J, {Dumbovi{\'c}} M, {Wimmer-Schweingruber} RF, {Temmer} M, {Lohf} H,
  {Wang} Y, et~al.
\newblock {Modeling the Evolution and Propagation of 10 September 2017 CMEs and
  SEPs Arriving at Mars Constrained by Remote Sensing and In Situ Measurement}.
\newblock {\em Space~Weather\/} {\bf 16} (2018) 1156--1169.
\newblock \doi{10.1029/2018SW001973}.

\bibitem[{{M{\"o}stl} et~al.(2010){M{\"o}stl}, {Temmer}, {Rollett}, {Farrugia},
  {Liu}, {Veronig} et~al.}]{mostl10}
{M{\"o}stl} C, {Temmer} M, {Rollett} T, {Farrugia} CJ, {Liu} Y, {Veronig} AM,
  et~al.
\newblock {STEREO and Wind observations of a fast ICME flank triggering a
  prolonged geomagnetic storm on 5-7 April 2010}.
\newblock {\em Geophys.~Res.~Lett.\/} {\bf 37} (2010) L24103.
\newblock \doi{10.1029/2010GL045175}.

\bibitem[{{Wood} et~al.(2011){Wood}, {Wu}, {Howard}, {Socker}, and
  {Rouillard}}]{wood11}
{Wood} BE, {Wu} CC, {Howard} RA, {Socker} DG, {Rouillard} AP.
\newblock {Empirical Reconstruction and Numerical Modeling of the First
  Geoeffective Coronal Mass Ejection of Solar Cycle 24}.
\newblock {\em Astrophys.~J.\/} {\bf 729} (2011) 70.
\newblock \doi{10.1088/0004-637X/729/1/70}.

\bibitem[{{Rodari} et~al.(2018){Rodari}, {Dumbovi{\'c}}, {Temmer},
  {Holzknecht}, and {Veronig}}]{rodari18}
{Rodari} M, {Dumbovi{\'c}} M, {Temmer} M, {Holzknecht} L, {Veronig} A.
\newblock {3D reconstruction and interplanetary expansion of the 2010 April
  $3^\mathrm{rd}$ CME}.
\newblock {\em Central European Astrophysical Bulletin\/} {\bf 42} (2018) 11.

\bibitem[{{Thernisien} et~al.(2006){Thernisien}, {Howard}, and
  {Vourlidas}}]{thernisien06}
{Thernisien} AFR, {Howard} RA, {Vourlidas} A.
\newblock {Modeling of Flux Rope Coronal Mass Ejections}.
\newblock {\em Astrophys.~J.\/} {\bf 652} (2006) 763--773.
\newblock \doi{10.1086/508254}.

\bibitem[{{Thernisien} et~al.(2009){Thernisien}, {Vourlidas}, and
  {Howard}}]{thernisien09}
{Thernisien} A, {Vourlidas} A, {Howard} RA.
\newblock {Forward Modeling of Coronal Mass Ejections Using STEREO/SECCHI
  Data}.
\newblock {\em Solar~Phys.\/} {\bf 256} (2009) 111--130.
\newblock \doi{10.1007/s11207-009-9346-5}.

\bibitem[{{Thernisien}(2011)}]{thernisien11}
{Thernisien} A.
\newblock {Implementation of the Graduated Cylindrical Shell Model for the
  Three-dimensional Reconstruction of Coronal Mass Ejections}.
\newblock {\em Astrophys.~J.~Suppl.\/} {\bf 194} (2011) 33.
\newblock \doi{10.1088/0067-0049/194/2/33}.

\bibitem[{{Howard} et~al.(2008){Howard}, {Moses}, {Vourlidas}, {Newmark},
  {Socker}, {Plunkett} et~al.}]{howard08}
{Howard} RA, {Moses} JD, {Vourlidas} A, {Newmark} JS, {Socker} DG, {Plunkett}
  SP, et~al.
\newblock {Sun Earth Connection Coronal and Heliospheric Investigation
  (SECCHI)}.
\newblock {\em Space~Sci.~Rev.\/} {\bf 136} (2008) 67--115.
\newblock \doi{10.1007/s11214-008-9341-4}.

\bibitem[{{Brueckner} et~al.(1995){Brueckner}, {Howard}, {Koomen}, {Korendyke},
  {Michels}, {Moses} et~al.}]{brueckner95}
{Brueckner} GE, {Howard} RA, {Koomen} MJ, {Korendyke} CM, {Michels} DJ, {Moses}
  JD, et~al.
\newblock {The Large Angle Spectroscopic Coronagraph (LASCO)}.
\newblock {\em Solar~Phys.\/} {\bf 162} (1995) 357--402.
\newblock \doi{10.1007/BF00733434}.

\bibitem[{{Wood} et~al.(2017){Wood}, {Wu}, {Lepping}, {Nieves-Chinchilla},
  {Howard}, {Linton} et~al.}]{wood17}
{Wood} BE, {Wu} CC, {Lepping} RP, {Nieves-Chinchilla} T, {Howard} RA, {Linton}
  MG, et~al.
\newblock {A STEREO Survey of Magnetic Cloud Coronal Mass Ejections Observed at
  Earth in 2008-2012}.
\newblock {\em Astrophys.~J.~Suppl.\/} {\bf 229} (2017) 29.
\newblock \doi{10.3847/1538-4365/229/2/29}.

\bibitem[{{Millward} et~al.(2013){Millward}, {Biesecker}, {Pizzo}, and {de
  Koning}}]{millward13}
{Millward} G, {Biesecker} D, {Pizzo} V, {de Koning} CA.
\newblock {An operational software tool for the analysis of coronagraph images:
  Determining CME parameters for input into the WSA-Enlil heliospheric model}.
\newblock {\em Space~Weather\/} {\bf 11} (2013) 57--68.
\newblock \doi{10.1002/swe.20024}.

\bibitem[{{Richardson} and {Cane}(2010)}]{richardson10}
{Richardson} IG, {Cane} HV.
\newblock {Near-Earth Interplanetary Coronal Mass Ejections During Solar Cycle
  23 (1996 - 2009): Catalog and Summary of Properties}.
\newblock {\em Solar~Phys.\/} {\bf 264} (2010) 189--237.
\newblock \doi{10.1007/s11207-010-9568-6}.

\bibitem[{{Mas{\'{\i}}as-Meza} et~al.(2016){Mas{\'{\i}}as-Meza}, {Dasso},
  {D{\'e}moulin}, {Rodriguez}, and {Janvier}}]{masias-meza16}
{Mas{\'{\i}}as-Meza} JJ, {Dasso} S, {D{\'e}moulin} P, {Rodriguez} L, {Janvier}
  M.
\newblock {Superposed epoch study of ICME sub-structures near Earth and their
  effects on Galactic cosmic rays}.
\newblock {\em Astron.~Astrophys.\/} {\bf 592} (2016) A118.
\newblock \doi{10.1051/0004-6361/201628571}.

\bibitem[{{Cargill} et~al.(1994){Cargill}, {Chen}, {Spicer}, and
  {Zalesak}}]{cargill94}
{Cargill} PJ, {Chen} J, {Spicer} DS, {Zalesak} ST.
\newblock {The deformation of flux tubes in the solar wind with applications to
  the structure of magnetic clouds and CMEs}.
\newblock {Hunt} JJ, editor, {\em Solar Dynamic Phenomena and Solar Wind
  Consequences, the Third SOHO Workshop\/} (1994), {\em ESA Special
  Publication\/}, vol. 373, 291.

\bibitem[{{Neugebauer}(2013)}]{neugebauer13}
{Neugebauer} M.
\newblock {Propagating Shocks}.
\newblock {\em Space~Sci.~Rev.\/} {\bf 176} (2013) 125--132.
\newblock \doi{10.1007/s11214-010-9707-2}.

\end{thebibliography}

\end{document}